\documentclass{aa}
\usepackage{txfonts}
\usepackage{graphicx}
\begin{document}
\title{The dark matter halos  of the bluest low surface brightness galaxies 
\thanks{Based on observations made with ESO Telescopes at Paranal under programme 69.B-0716.}}
\markboth{E. Zackrisson et al.: }{The dark matter halos of the bluest LSBGs} 
\authorrunning{E. Zackrisson et al.}
\titlerunning{The dark matter halos of the bluest LSBGs} 
\author{E. Zackrisson\inst{1,2} \and N. Bergvall\inst{1} \and T. Marquart\inst{1} \and G. \"Ostlin\inst{3}}
\offprints{Erik Zackrisson, 
\email{ez@astro.uu.se}}
\institute{Department of Astronomy and Space Physics, Box 515, S-75120 Uppsala, Sweden \and
Tuorla Observatory, V\"ais\"al\"antie 20, FI-21500 Piikki\"o, Finland
\and  Stockholm Observatory, AlbaNova University Center, 106 91 Stockholm, Sweden}
\date{Received 13 September 2005 / Accepted 6 March 2006}
\abstract{We present $BVI$ photometry and long-slit H$\alpha$ rotation curve data obtained with ESO VLT/FORS2 for six low surface brightness galaxies with extremely blue colours and very faint central regions. We find no evidence for a steep central density cusp of the type predicted by many N-body simulations of cold dark matter (CDM) halos. Our observations are instead consistent with dark matter halos characterized by cores of roughly constant density, in agreement with previous investigations. While unremarkable in terms of the central density slope, these galaxies appear very challenging for existing CDM halo models in terms of average central halo density, as measured by the $\Delta_{V/2}$ parameter. Since most of our target galaxies are bulgeless disks, our observations also disfavour a recently suggested mechanism for lowering the central mass concentration of the halo by means of a fast collapse phase, as this scenario predicts that the original CDM profile should still be detectable in bulgeless galaxies. Other potential ways of reconciling the CDM predictions with these observations are discussed.
}
\maketitle

\section{Introduction} 
The cold dark matter (CDM) scenario, in which the dark matter particles are assumed to be non-relativistic at the time of decoupling, and to interact predominantly through gravity, has been very successful in explaining the formation of large-scale structures in the Universe (e.g. Primack \cite{Primack}). On the scales of galaxies, the CDM predictions of halo shapes, substructure and density profiles have however not yet been confirmed in any convincing way. 

Dwarf galaxies and low surface brightness galaxies (LSBGs) are believed to be more or less completely dominated by dark matter, thereby making them among the best probes of the density profiles of dark matter halos on galactic scales. A lot of recent research in this field has revolved around apparent discrepancies between the CDM halo density profiles predicted by N-body simulations (e.g. Navarro et al. \cite{Navarro et al. 96}, \cite{Navarro et al. 97}; hereafter NFW) and the dark halo density profiles inferred from observations (see e.g. de Blok et al. \cite{de Blok et al. a}; de Blok \& Bosma \cite{de Blok & Bosma}). Both the observed slope of the innermost density profile (the core/cusp problem) and the overall shape of LSBG rotation curves  are in conflict with the N-body results, indicating a potentially serious problem for the CDM scenario.

To remedy this situation, numerous solutions have been proposed. Some of the discrepancies may be reduced by assuming that dark matter is not cold, but rather self-interacting (Spergel \& Steinhardt \cite{Spergel & Steinhardt}), warm (Bode et al. \cite{Bode et al.}), annihilating (Kaplinghat et al. \cite{Kaplinghat et al.}), decaying (Cen \cite{Cen}), or that the dark energy of the Universe has a phantom-like equation of state (Kuhlen et al. \cite{Kuhlen et al.}). Another option may be to drop the notion of dark matter altogether and instead modify the laws of gravity (e.g. McGaugh \& de Blok \cite{McGaugh & de Blok b}). Most observational studies so far have however assumed the dark matter halos to be spherical, whereas CDM in fact predicts triaxial halos. It has been suggested that a realistic treatment of disk dynamics inside these more complicated potentials may possibly remove the discrepancy (Hayashi et al. \cite{Hayashi et al. b}). Yet another possibility is that the dark matter domination of some target galaxies have been overestimated (Graham \cite{Graham}; Fuchs \cite{Fuchs}) and that some complicated baryonic process may have significantly affected the density profile.   

Despite a large number of failures to confirm the NFW predictions for the CDM halo profile on galactic scales, there are at least two objects for which the data have been reported to be of sufficient quality to allow well-constrained fits, yet NFW profiles still appear reasonably consistent with the observations: NGC 5963 (Simon et al. \cite{Simon et al.}) and DDO 9 (de Blok \cite{de Blok b}). Is it then possible that some galaxies have NFW halos whereas others do not? Could it be that the apparent discrepancy between predicted and observed halos is due to some bias related to how the target galaxies have been selected?  
 
Here, we investigate the central kinematics of a class of LSBGs not previously targeted by similar investigations. Our targets are bulgeless disks with extremely blue colours and very low surface brightness centres; properties which could possibly imply more dark-matter dominated central regions and alleviate potential problems of baryon-domination. In Sect. 2 we describe the observations and the selection criteria of our galaxy sample. In Sect. 3 the rotation curves are presented and in Sect. 4 the luminosity profiles. Sects. 5 and 6 compares the observed average central density and the inner slope of the density profile to the CDM predictions. Sect. 7 discusses the implications of our results and a number of potential problems with our analysis. In Sect. 8, our findings are summarized.
\begin{figure}[t]
\resizebox{\hsize}{!}{\includegraphics{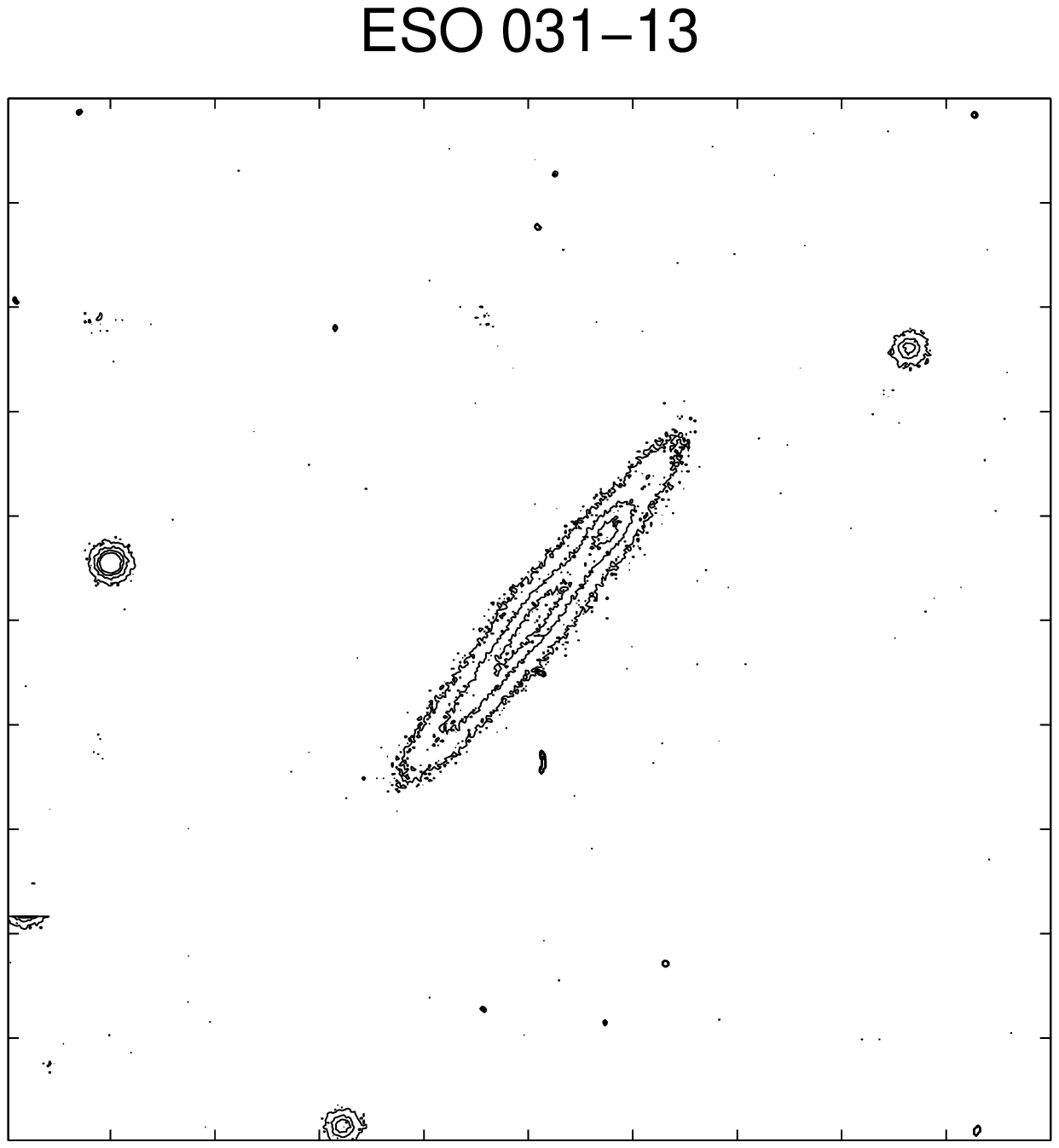}\includegraphics{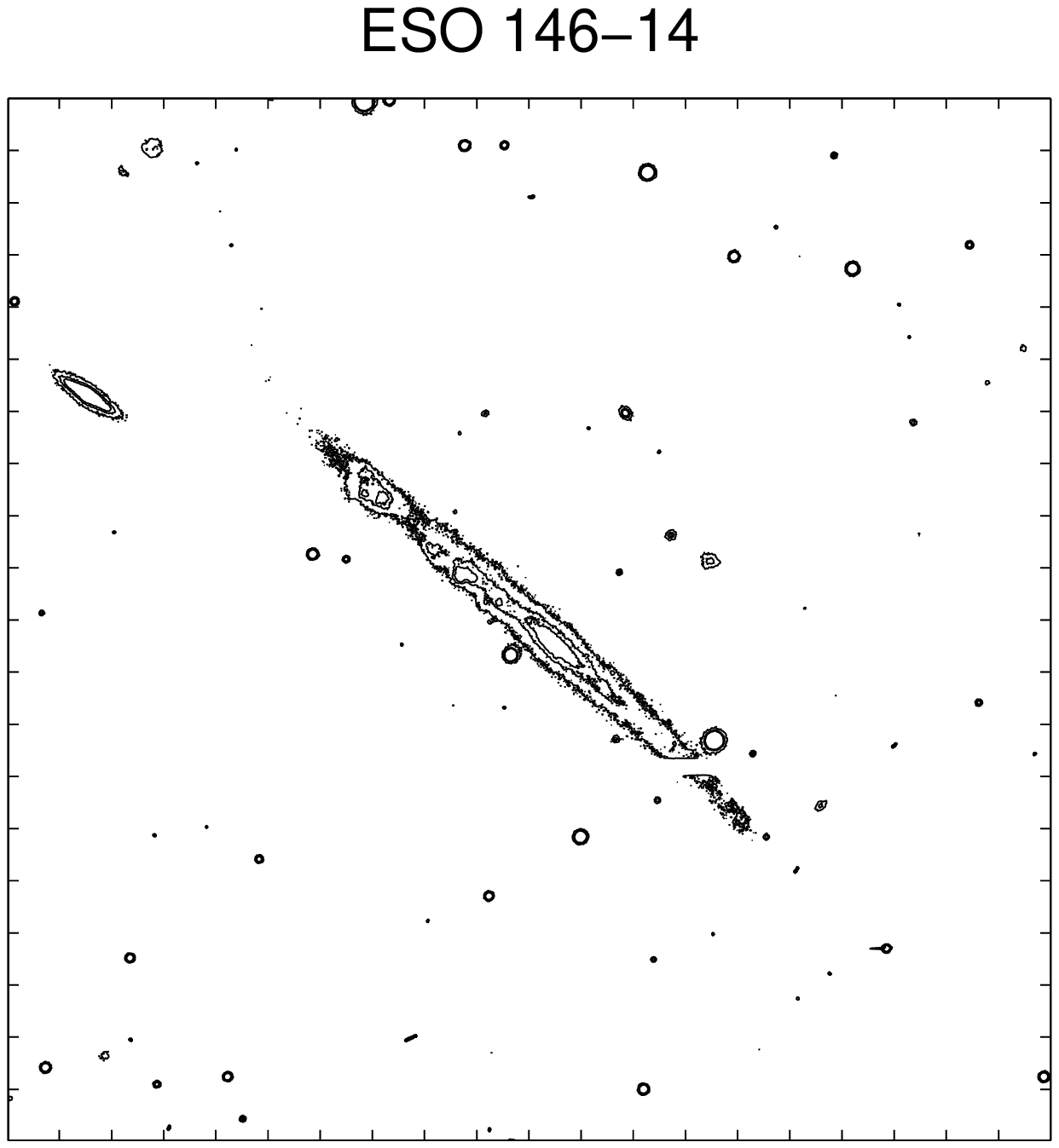}\includegraphics{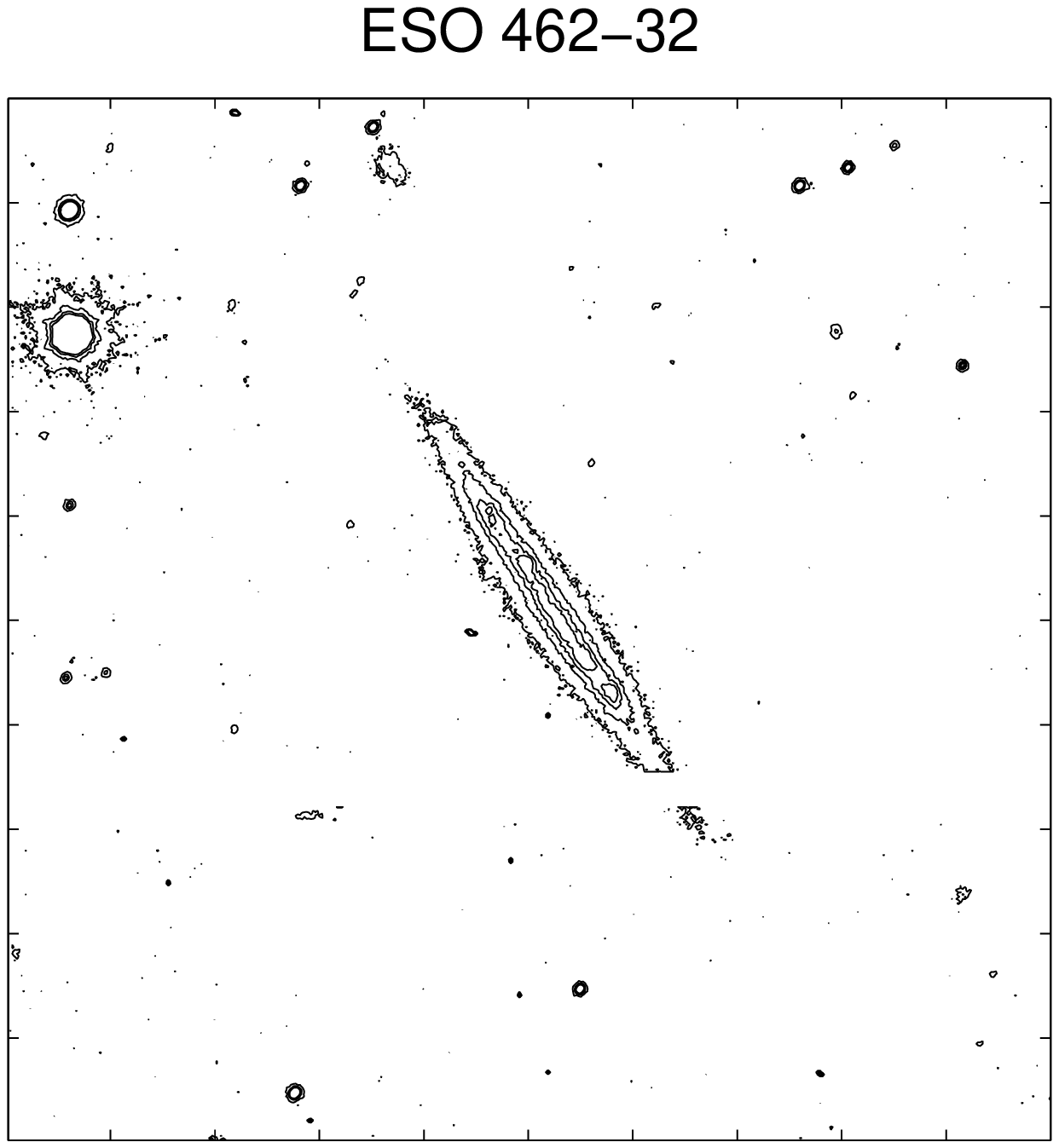}}
\resizebox{\hsize}{!}{\includegraphics{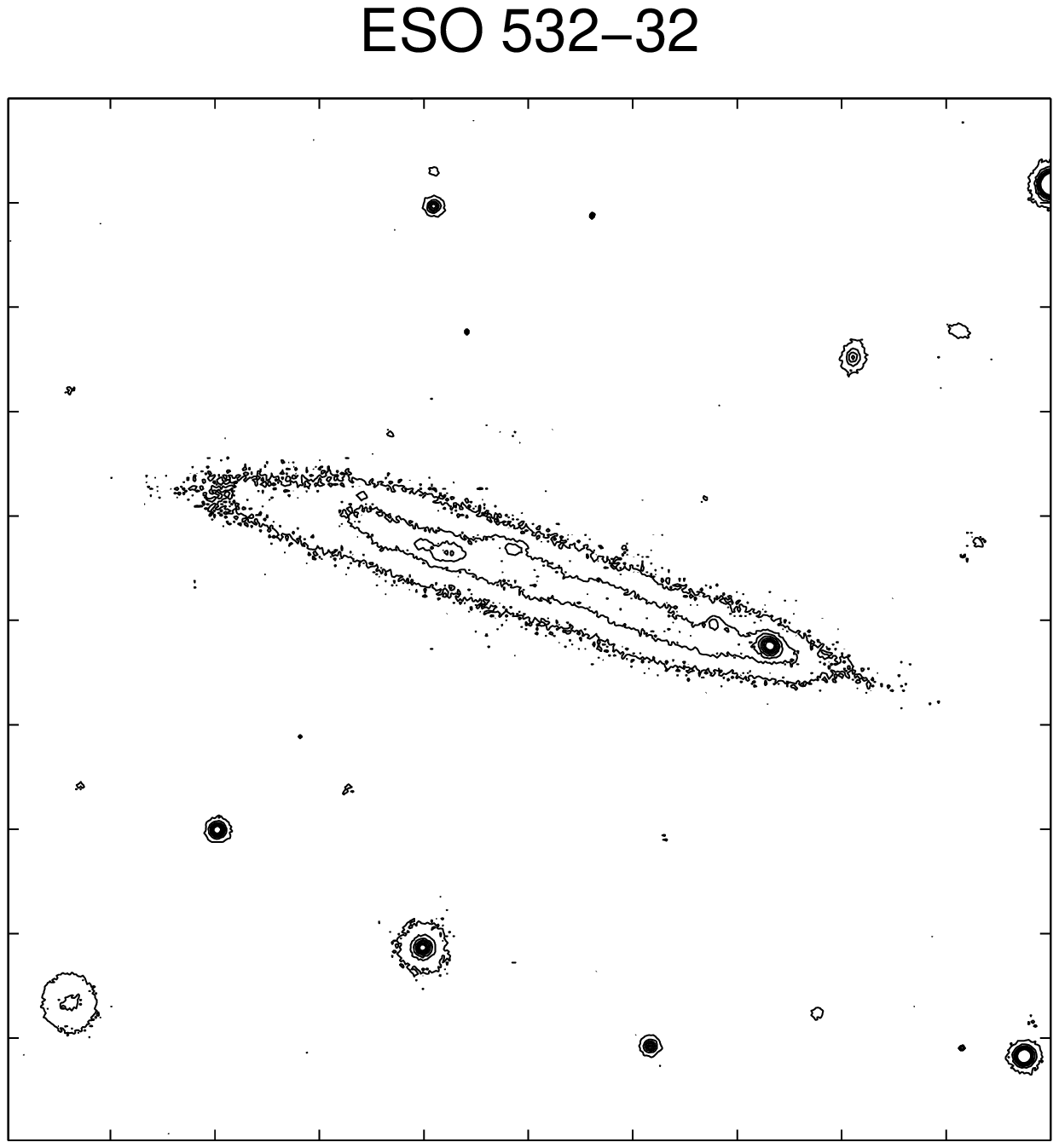}\includegraphics{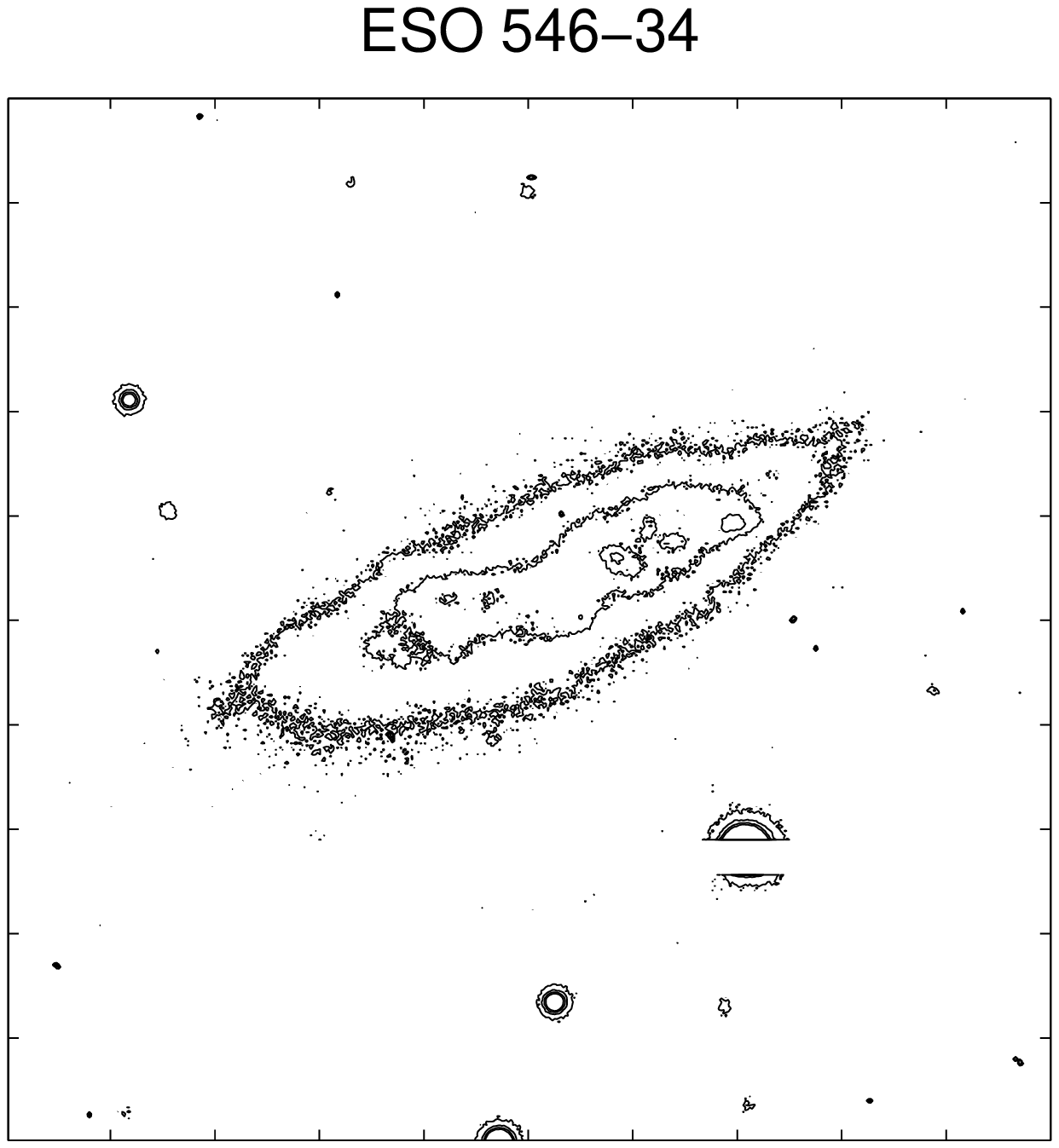}\includegraphics{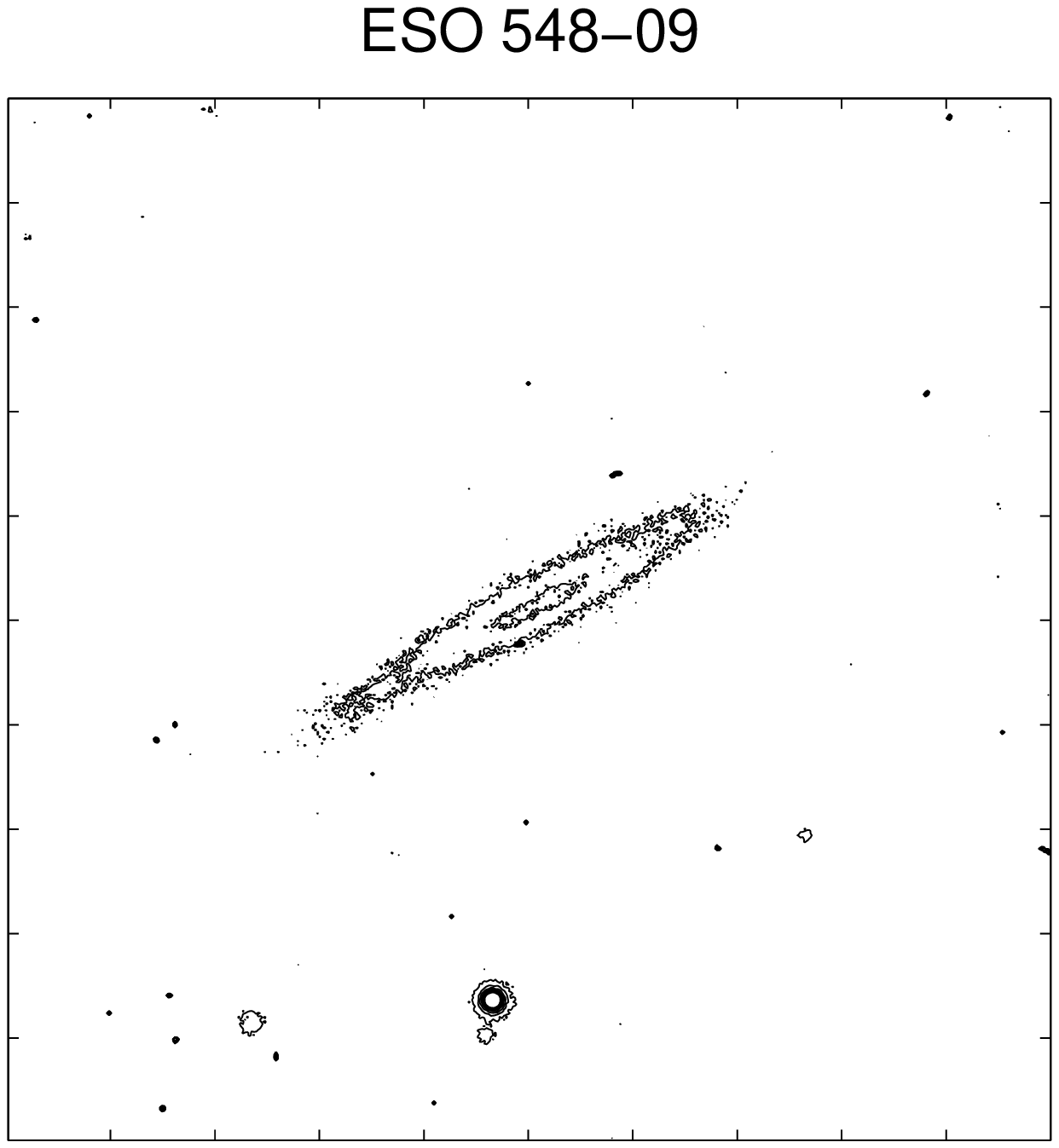}}
\caption[]{Isophotal B-band images of the target galaxies. The field size is $2\arcmin \times 2\arcmin$, except for ESO 146-14, for which it is $4\arcmin \times 4\arcmin$. The outermost isophotes plotted correspond to $\mu_B=24.5\pm0.2$ mag arcsec$^{-2}$, and the flux steps between the brighter contours are constant on a linear scale (but different from galaxy to galaxy, to clearly display the structures present). North is up and east is to the left. The horizontal line visible in several of the images is caused by the gap between the two FORS2 CCDs .}
\label{Thumbnails}
\end{figure}

\section{Observations, data reduction and sample properties}
\subsection{Selection criteria}
The six disk-like LSBGs used in this investigation were selected from the ESO-Uppsala catalogue (Lauberts \& Valentijn \cite{lauberts}) with the criteria that the targets should have:

1) An average $B-R\lesssim0.5$ mag inside the region with $B$-band surface brightness $20.5 \leq \mu_B \ (\mathrm{mag \ arcsec^{-2}})\leq 26$; 

2) A $B$-band surface brightness inside the central $5\arcsec$ of $\mu_{B,0}\gtrsim 22.3$ mag arcsec$^{-2}$; 

3) A high inclination, but be separated from edge-on projection by at least $5^\circ$ (under the approximation of an infinitely thin disk).

The first and second selection criteria are similar to those used in our previous studies of extremely blue LSBGs (e.g. R\"onnback \& Bergvall \cite{Rönnback & Bergvall a}; Bergvall et al. \cite{Bergvall et al. a}; Zackrisson et al. \cite{Zackrisson et al.}), which indicate that objects of this type are metal-poor and almost extinction-free. The galaxies selected are \object{ESO 031-13}, \object{ESO 146-14}, \object{ESO 462-32}, \object{ESO 532-32}, \object{ESO 546-34} and \object{ESO 548-09}. Out of these, ESO 146-14, 546-34 and 548-09 have featured in our previous papers on blue LSBGs.
As far as we can tell, LSBGs of this type have not been targeted by previous dark halo investigations using optical rotation curve data (e.g. de Blok et al. \cite{de Blok et al. a}; de Blok \& Bosma \cite{de Blok & Bosma}; Swaters et al. \cite{Swaters et al.}). The selection criteria are further discussed in Sect. 2.3.

\subsection{Observations and reductions}
\begin{table}[t]
\caption[]{List of objects, integration times and observation dates. Photometric observations are labeled $B$,$V$ or $I$ depending on filter. Spectroscopic long-slit observations are labeled = and $\|$ for slits aligned along the major and minor axis of the galaxy, respectively.}
\begin{flushleft}
\begin{tabular} {lll}
\hline
\hline
ESO id.  & Integration time (s) & Date (yymmdd)\\
\noalign{\smallskip}
\hline
\noalign{\smallskip}
031-13  & $B$(852) & 020801, 020813\\
	 & $V$(372) & 020801, 020813\\
	 & $I$(1040) & 020801, 020813\\
	 & =(6324) & 020913\\
	 & $\|$(1563) & 020831\\
\\
146-14  & $B$(120) & 020706\\
	 & $V$(68) & 020706\\
	 & $I$(120) & 020706\\
	 & =(2664) & 020718\\
\\
462-32  & $B$(320) & 020506\\
	 & $V$(136) & 020506\\
	 & $I$(372) & 020506\\
	 & =(6324) & 020518\\
	 & $\|$(1560) & 020518\\
\\
532-32  & $B$(270) & 020706\\
	 & $V$(120) & 020706\\
	 & $I$(304) & 020706\\
	 & =(3162) & 020714\\
	 & $\|$(1560) & 020706\\
\\
546-34  & $B$(744) & 020711, 020801\\
	 & $V$(340) & 020711, 020801\\
	 & $I$(880) & 020711, 020801\\
	 & =(6324) & 020911\\
	 & $\|$(2080) & 020815\\
\\
548-09  & $B$(542) & 020801\\
	 & $V$(256) & 020801\\
	 & $I$(744) & 020801\\
	 & = (6324) & 021005\\
	 & $\|$(2080) & 021005\\
\noalign{\smallskip}
\hline
\end{tabular} \\
\end{flushleft}
\label{Obstable1}
\end{table}
All observations were carried out in service mode with the FORS2 instrument at the VLT-UT4 Yepun 8.2m ESO telescope in 2002. Imaging in $BVI$ was carried out at a seeing of $\leq1.2\arcsec$, while long-slit spectroscopy along both the major and minor axis of the target galaxies was carried out at a seeing of  $\leq0.8\arcsec$. The spectroscopy was performed using the $1.0\arcsec$ slit and the 1200R grism, with a wavelength range of 5750--7310 \AA{ }and a spectral resolution of 35 km s$^{-1}$ per pixel at the central wavelength. Contour plots of the $B$-band images are displayed in Fig.~\ref{Thumbnails}.
\begin{table*}[t]
\caption[ ]{Target galaxy data. Here, $m_B$ and $M_B$ represent the apparent and absolute integrated $B$-band magnitudes inside the $\mu_{B,0}$ = 26.5 mag arcsec$^{-2}$ isophote. $B-V$ and $V-I$ denote the integrated colours inside the same radius, whereas $<$$B-R$$>$ (taken from the ESO-Uppsala catalogue; Lauberts \& Valentijn \cite{lauberts}) is the average colour inside the $20.5 \leq \mu_B\ (\mathrm{mag \ arcsec^{-2}})\leq 26$ region. $h_I$ represents the $I$-band scale length derived from the outer part of the surface brightness profile. Both the true $B$-band central surface brightness, $\mu_{B,0}$, and the $B$-band surface brightness of an exponential profile extrapolated to the centre, $\mu^\mathrm{D}_{B,0}$, have been corrected for inclination, $i$, assuming an infinitely thin disk. $v_\mathrm{sys, hel}$ denotes the systemic, heliocentric velocities and $D$ the estimated distances.  All magnitudes have been corrected for galactic extinction (Schlegel et al. \cite{Schlegel et al.}).}
\begin{flushleft}
\begin{tabular} {llllllllllllllll}
\noalign{\smallskip}
\hline
\hline
\noalign{\smallskip}
ESO id.  & $v_\mathrm{sys, hel}$ & $D$  & m$_B$  & M$_B$  & $\mu_{B,0}$ & $\mu^\mathrm{D}_{B,0}$ & $B-V$ & $<$$B-R$$>$ & $V-I$ & $h_I$ & $i$ &\\
& (km s$^{-1})$ & (Mpc) & (mag) & (mag) & (mag & (mag & (mag) & (mag) & (mag) & (kpc) & ($^\circ$)&\\
&  &  & & & arcsec$^{-2}$) & arcsec$^{-2}$) &  &  &\\
\noalign{\smallskip}
\hline
\noalign{\smallskip}
031-13 & 6401 & 82.7 & 16.7 & -17.9 & 24.3 & 23.2 & 0.49 & 0.28 & 0.95 & 3.5 & 83\\ 
146-14 & 1686 & 21.0 & 14.9 & -16.7 & 23.5 & 24.0 & 0.21 & 0.25 & 0.70 & 2.8 & 82\\
462-32 & 2848 & 38.7 & 16.7 & -16.3 & 23.8 & 23.2 & 0.45 & 0.38 & 1.31 & 1.5 & 83\\ 
532-32 & 2674 & 35.8 & 16.3 & -16.4 & 24.4 & 23.4 & 0.42 & 0.48 & 0.60 & 2.4 & 79\\ 
546-34 & 1582 & 19.5 & 15.5 & -15.9 & 24.2 & 22.8 & 0.21 & 0.41 & 0.54 & 1.3 & 76\\ 
548-09 & 1832 & 22.7 & 17.3 & -14.4 & 24.7 & 24.4 & 0.45 & 0.38 & 0.80 & 1.5 & 80\\ 
\noalign{\smallskip}
\hline
\end{tabular} \\
\end{flushleft}
\label{Obstable2}
\end{table*}

Observing dates and exposure times are summarized in Table.~\ref{Obstable1}. Due to technical problems, the minor-axis observations for ESO 146-14 were unfortunately not properly executed. The pipeline-reduced images were recalibrated using standard stars, stacked and sky subtracted using the ESO-MIDAS package. Spectra were reduced and wavelength-calibrated from arcspectra, using the same software. After stacking, all spectra were recalibrated to spatial bins of $0.75 \arcsec$, except for the major-axis data of ESO 532-32, where the low signal-to-noise required the use of bins $1.25 \arcsec$ wide. 

Galaxy centres and inclinations were determined from the outer isophotes of the images, the latter under the assumption of an infinitely thin disk. Surface brightness profiles were derived from in-house software developed exclusively for this purpose. Distances to all galaxies were estimated from the systemic, heliocentric redshifts after converting to the centroid of the Local Group using IAU specifications and correcting for Virgocentric infall using the model by Schechter (\cite{Schechter}). A distance to the Virgo cluster of 17 Mpc, corresponding to a Hubble constant of $H_0=75 \ \mathrm{km\ s^{-1}\ Mpc^{-1}}$, was assumed.  All magnitudes were corrected for Galactic extinction using the $B$-band extinction maps by Schlegel et al. (\cite{Schlegel et al.}), and the standard ($\mathrm{R}_V=3.1$) extinction law (Cardelli et al. \cite{Cardelli et al.}), as implemented in the NED. Due to the different extinction corrections used here, the data for ESO 146-14 and ESO 546-34 differ slightly from those presented in Zackrisson et al. (\cite{Zackrisson et al.}).

Table \ref{Obstable2} summarizes the $BVI$ magnitudes, heliocentric systemic velocities, distances, disk scale lengths, inclinations and central surface brightness levels of our objects. Here, $\mu_{B,0}$ refers to the true $B$-band central surface brightness (integrated over the central $1.2 \arcsec$, which corresponds to the maximum seeing disk), whereas $\mu^\mathrm{D}_{B,0}$ refers to the surface brightness of a fitted exponential disk extrapolated to the centre. The disk scale lengths are derived from the outer parts of the $I$-band surface brightness profile, as discussed in Sect. 4.

\subsection{General sample properties}
For measuring the dark halo density profile, it would appear advantageous to target the most dark-matter dominated galaxies, thereby minimizing the complicated effects associated with luminous baryons. Since the relation between galaxies and their dark halos is far from well-understood, this exercise is however not trivial.

The galaxies used in this investigation have been selected on the basis of faint central regions, high inclinations and very blue colours. What bearing may these selection criteria have on the dark matter properties of these objects and the accuracy with which they may be derived? 

In McGaugh \& de Blok (\cite{McGaugh & de Blok a}), a strong correlation was demonstrated between the dynamical mass-to-light ratio and the central disk surface brightness. This relation would imply that selecting the LSBGs with the faintest central regions should be advantageous for testing CDM models, unless the baryonic mass-to-light ratio is also unusually large for these objects. To target the most extreme low surface brightness galaxies, we have selected objects with a surface brightness inside the central $5\arcsec$ of $\mu_{0,B}>22.3$ mag arcsec$^{-2}$, in combination with a high inclination. After correcting for inclination, four out of our six objects turn out to have $\mu_{0,B}>24$ mag arcsec$^{-2}$, making them among the most low-surface brightness objects used so far for measuring the density profiles of dark halos.

Five out of our six targets (the exception being ESO 146-14) are underluminous in the central region compared to what would be expected from an exponential disk fitted to the outer isophotes. Hence, they appear to be completely bulgeless objects, which should make them ideal for testing the scenario advanced by Mo \& Mao (\cite{Mo & Mao}) to explain the discrepancy between observations and the halo profiles predicted by CDM. Mo \& Mao suggest that feedback associated with a fast collapse phase of baryons during the galaxy formation process could lower the high central densities predicted by CDM and simultaneously explain the presence of bulges. If this scenario is correct, the original CDM halo profiles should therefore still be observable in bulgeless disk galaxies dominated by dark matter, such as the ones presented here. 

Selecting galaxies with high inclinations comes with advantages as well as disadvantages. By targeting disk galaxies close to edge-on, it is possible to identify the most extreme LSBGs in surveys limited by surface brightness, and to minimize the inclination correction to line-of-sight velocities measured by long-slit spectroscopy, as well as the uncertainties in the position angle. High-inclination galaxies are however also sensitive to extinction and projection effects, especially in the central regions. The exact inclination limit at which  these problems become severe is unfortunately somewhat uncertain, as further discussed in Sect. 7.1. 

Since blue colours imply a lower stellar mass-to-light ratio $M/L_\star$ (e.g. Bell \& de Jong \cite{Bell & de Jong}), one may naively expect that blue LSBGs should be more dark-matter dominated than their red counterparts for a given central luminosity (surface brightness). This conjecture does however assume that blue and red galaxies are located inside identical dark matter halos, which is not necessarily the case. Graham (\cite{Graham}) examined the ratio of stellar disk to dynamical mass and did indeed find it to decrease for bluer galaxies. Zavala et al. (\cite{Zavala et al.}) on the other hand estimated the ratio of baryonic to dynamical mass and found a possible trend in the opposite direction. From an evolutionary point of view, the Zavala et al. result can be understood in a scenario in which the dark halos that formed early are more concentrated than those that formed late (Navarro et al. \cite{Navarro et al. 97}, Bullock et al. \cite{Bullock et al.}), and in which blue galaxies are younger than red ones in an absolute sense. Although the last condition may seem reasonable, it is difficult to prove for LSBGs at the present time (Zackrisson et al. \cite{Zackrisson et al.}), since the star formation histories may also differ between red and blue stellar populations. 

Although the relation between colour and dark matter properties is obscure at best, targeting very blue LSBGs does have other advantages. Since these objects are seen to be low-metallicity objects with little internal extinction (e.g. R\"onnback \& Bergvall \cite{Rönnback & Bergvall b}, Bergvall et al. \cite{Bergvall et al. a}), chemical evolution is less of a concern when estimating the $M/L_\star$ of their stellar populations (Zackrisson et al. \cite{Zackrisson et al.}). When retrieving the rotation curve from high-inclination disks, dust effects are also likely to be minimized. We may furthermore be confident that the central surface brightness levels of our targets have not been significantly underestimated due to dust effects.
\begin{figure}[t]
\resizebox{\hsize}{!}{\includegraphics{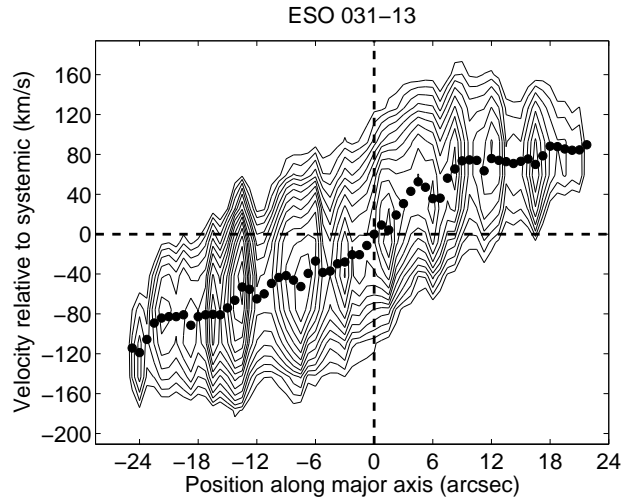}}
\caption[]{Position-velocity diagram for the H$\alpha$ data of ESO 031-13. The contours trace the H$\alpha$ flux on a logarithmic scale, whereas the dots indicate the rotational velocities resulting from a Gaussian fit to the line profile in each radial bin. The dashed vertical line indicates the adopted centre of the galaxy and the dashed horizontal line the systemic velocity. The line profiles do not display any obvious systematic tails towards the systemic velocity.}
 \label{posvelfig}
\end{figure}

\begin{figure*}[t]
\resizebox{\hsize}{!}{\includegraphics{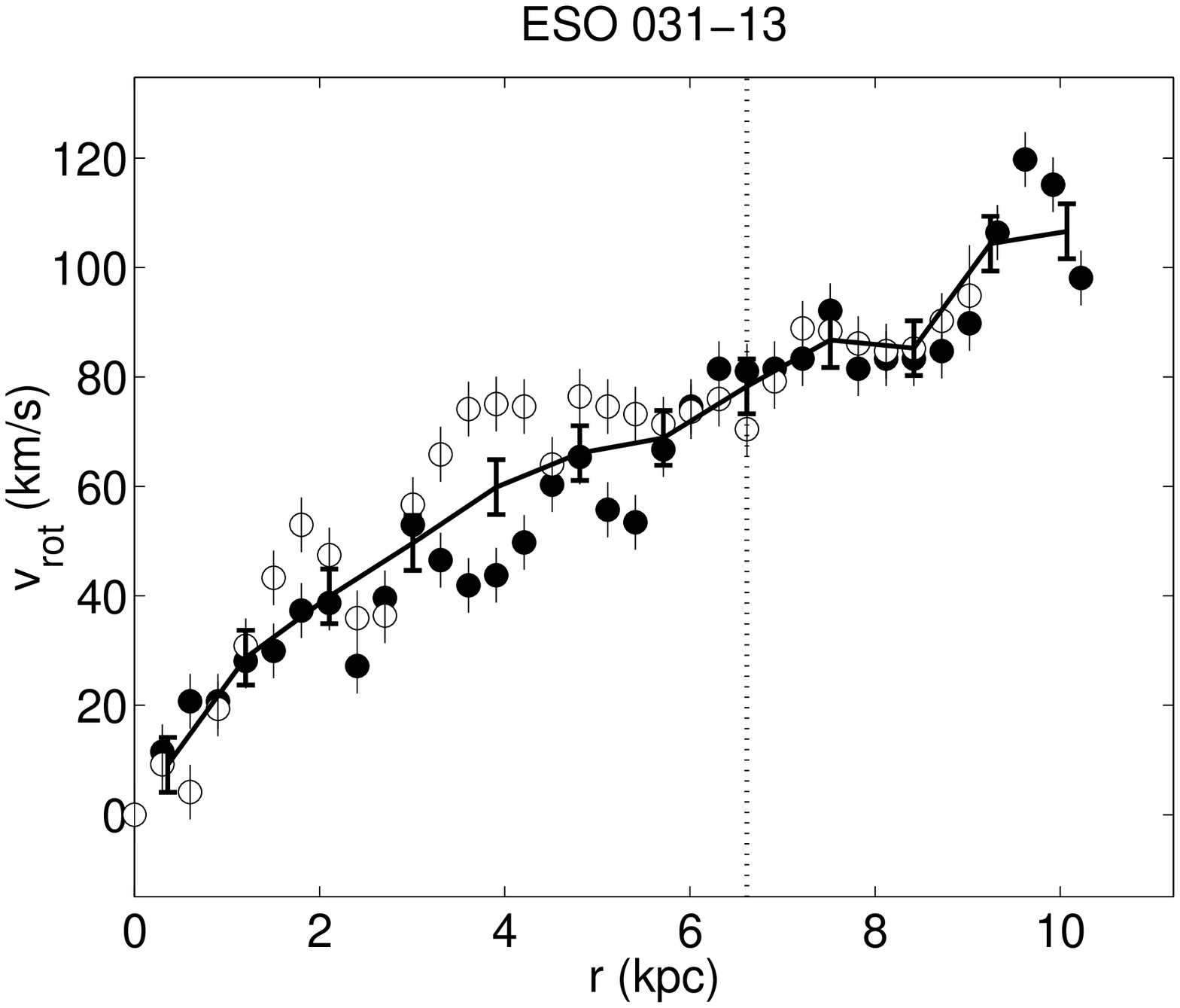}\includegraphics{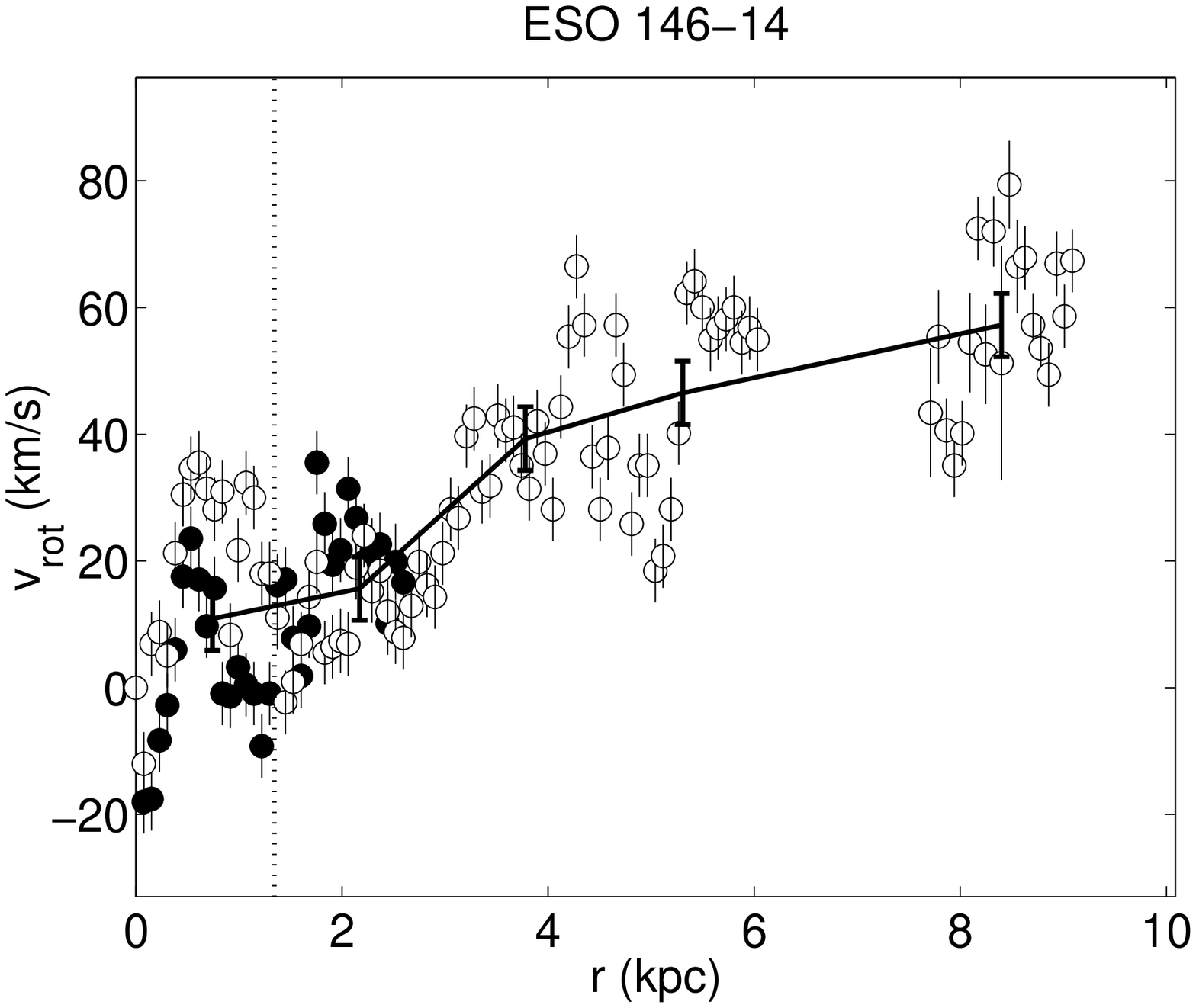}\includegraphics{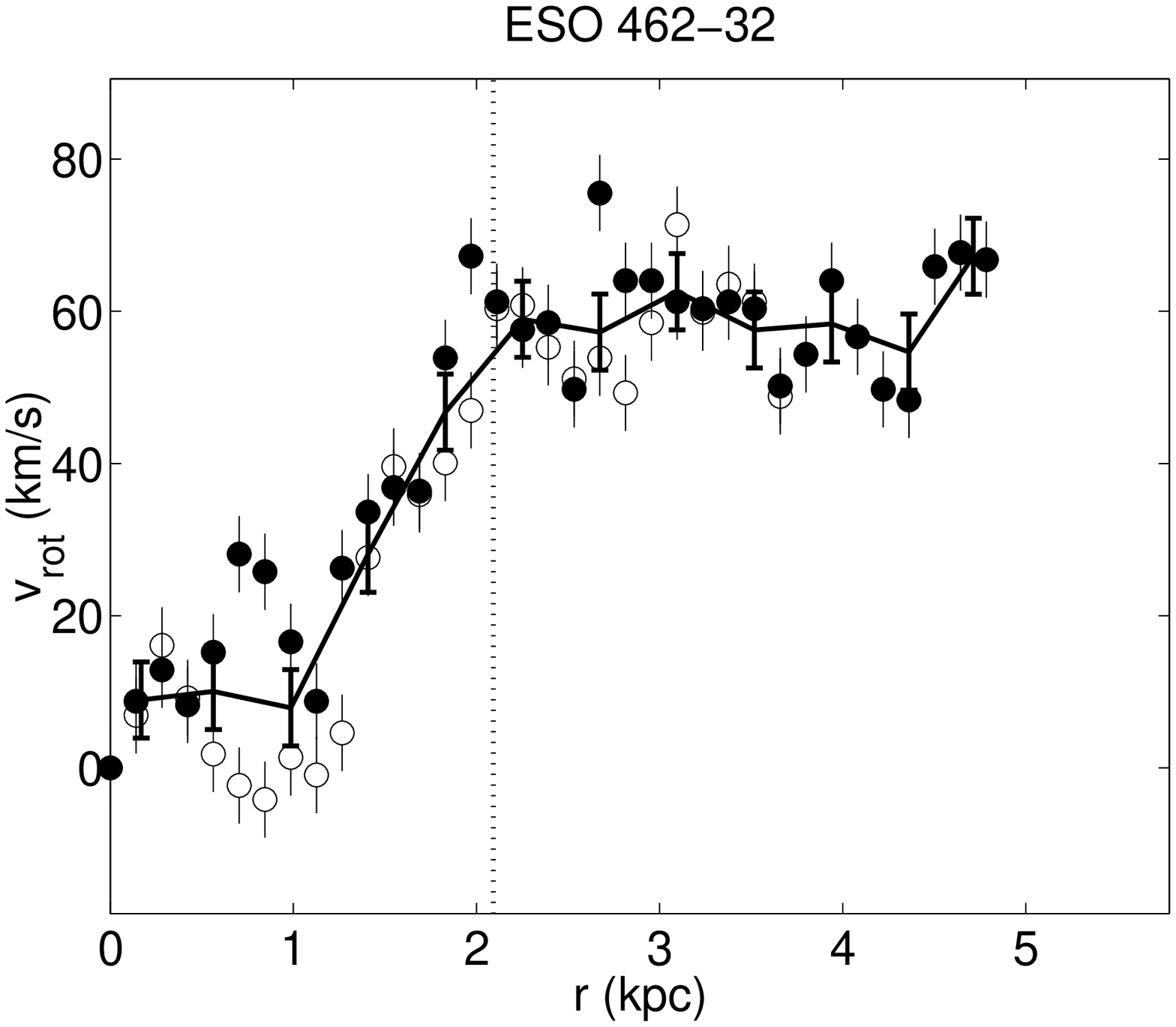}}
\resizebox{\hsize}{!}{\includegraphics{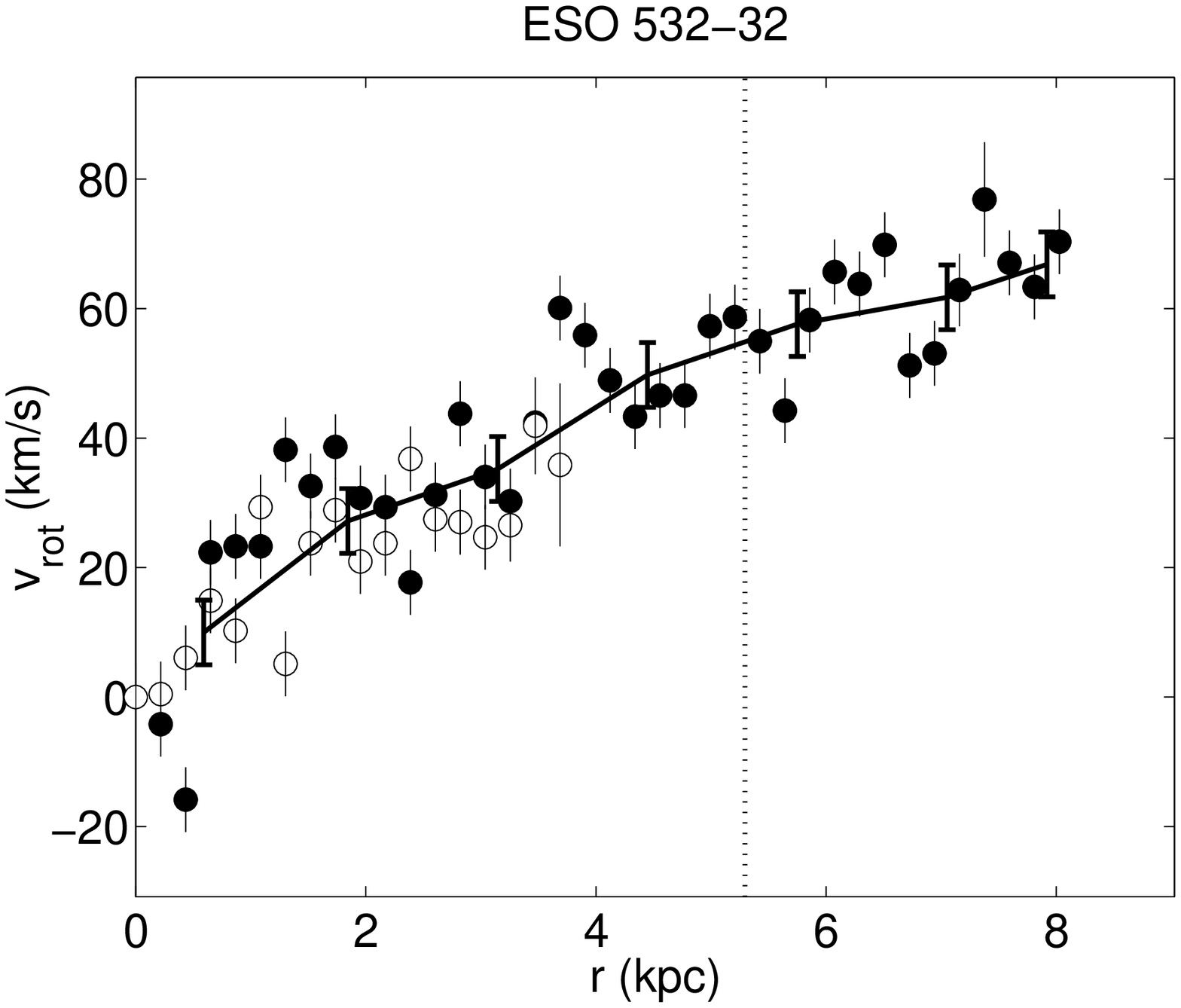}\includegraphics{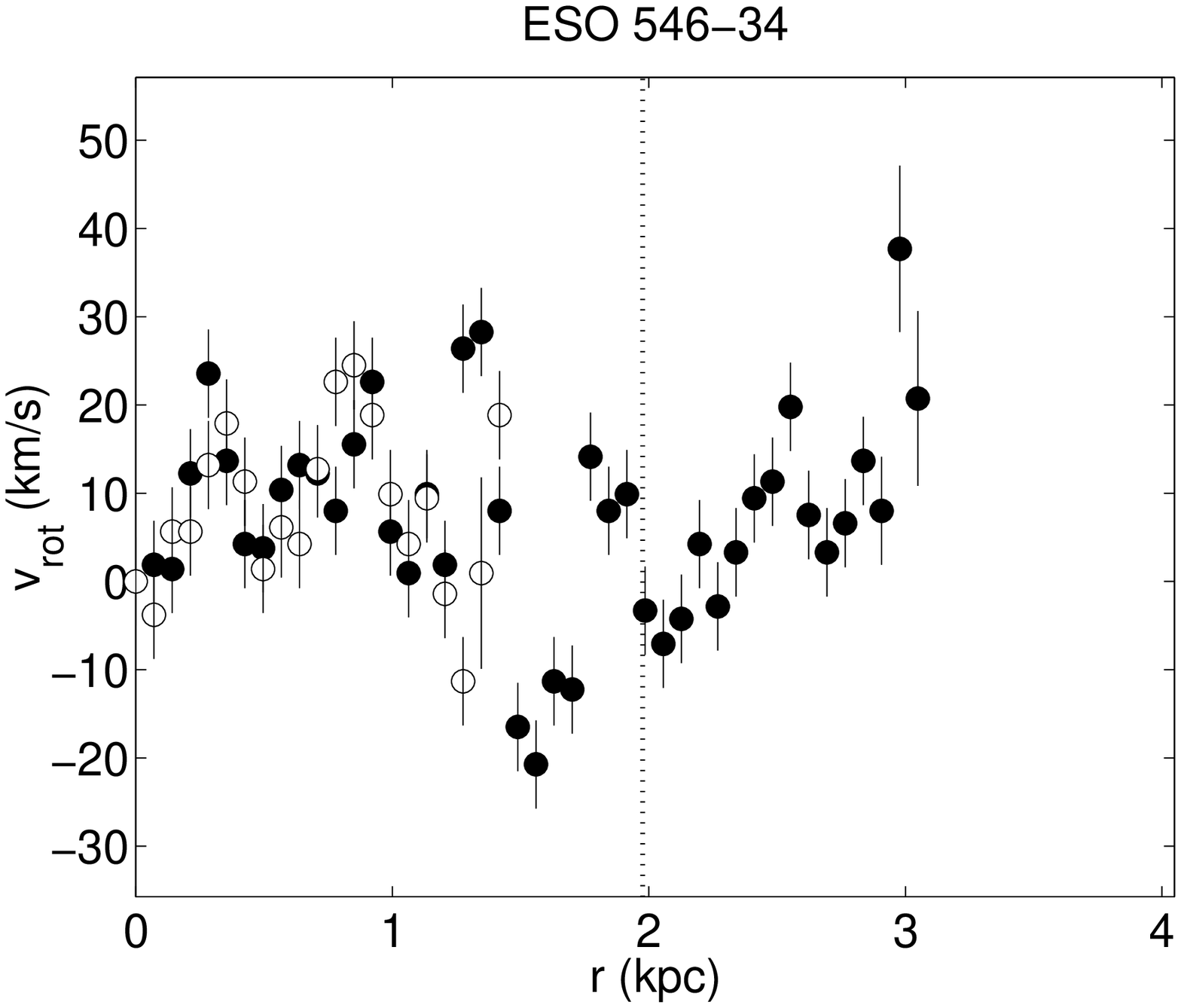}\includegraphics{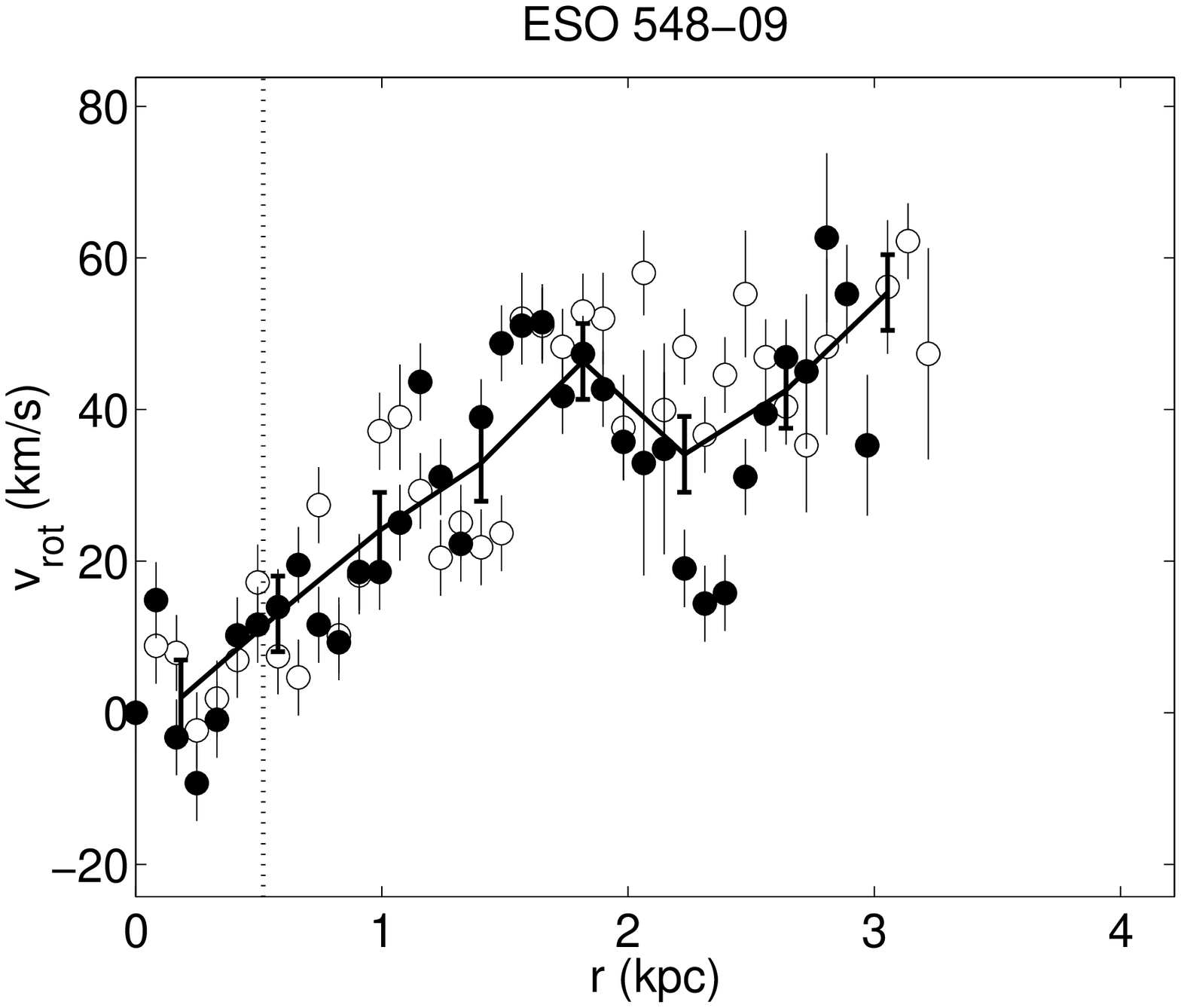}}
\caption[]{The major-axis rotation curves of the target galaxies. Black and white filled circles represent data from approaching and receding sides, respectively. The exception is ESO 546-34, for which the amplitude of the rotation curve is too low and the velocity field too disturbed to accurately determine this. For all other objects, symmetrised and spatially averaged rotation curves are plotted as black lines. For each galaxy, a dotted, vertical line indicates the radius at which the surface brightness profile starts to deviate from the exponential disk determined from the outer isophotes.}
\label{fig_RC_major}
\end{figure*}

\section{Rotation curves} 
Rotation curves have been derived along both minor and major axes by means of  fitting a gaussian to the line profile of the H$\alpha$ (6563 \AA) emission line. For edge-on disks, this fitting procedure may underestimate the true velocity because of projection effects, since a substantial amount of the light can originate at greater radii in the disk, where the line-of-sight velocities are lower. In these cases, more sophisticated fitting techniques should be employed to recover the true rotation velocity (e.g. Kregel \& van der Kruit \cite{Kregel & van der Kruit}; Gentile et al. \cite{Gentile et al. a}). Although the inclinations of our targets are  high and based on the approximation of an infinitely thin disc (therefore probably slightly underestimated), the emission-line profiles of these galaxies do not show any obvious systematic tail towards the systemic velocity, as would be expected if projection effects were significant. As an example, the H$\alpha$ position-velocity diagram for ESO 031-13 is displayed in Fig.~\ref{posvelfig}. The lack of systematic asymmetries in the line profiles leads us to assume that the disks are very thin and that kinematic projection effects are small. The possibility that a slight systematic skewness could be masked by the limited spectral resolution of these observations, in combination with a patchy distribution of emission-line regions in the disk, can however not be completely ruled out. See Sect. 7.1 for a more detailed discussion about potential projection effects. 

The approximate centre of each galaxy was estimated from the outer isophotes of the continuum radiation. Because of the flat light profile encountered in many cases (see Sect. 4), the exact centre was then determined by maximizing the symmetry between the two sides of the rotation curve inside a limited spatial region. In order to derive mass profiles of the galaxies, the major-axis rotation curve was folded and spatially averaged over bins large enough to produce a reasonably smooth appearance. No artificial smoothing (of the type used in e.g. de Blok et al. \cite{de Blok et al. a} and de Blok \& Bosma \cite{de Blok & Bosma}) has however been employed. The error bars on the spatially averaged rotation curves were simply taken to be the uncertainties in the mean velocity of each bin. A minimum  uncertainty of 5 km/s was finally imposed, to avoid unrealistically small errors associated with regions of the spectra with very high S/N.  

\begin{figure*}[t]
\resizebox{\hsize}{!}{\includegraphics{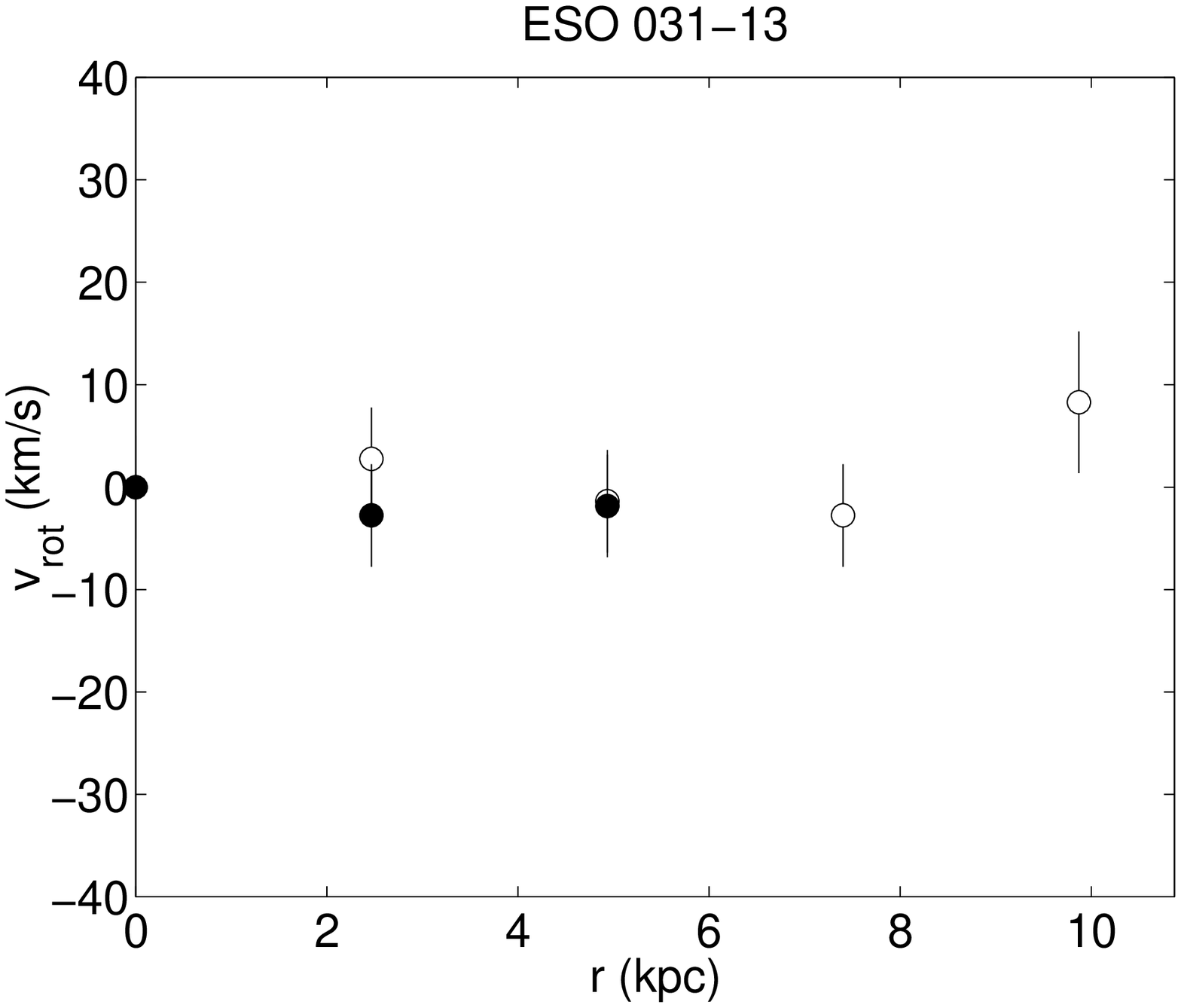}\includegraphics{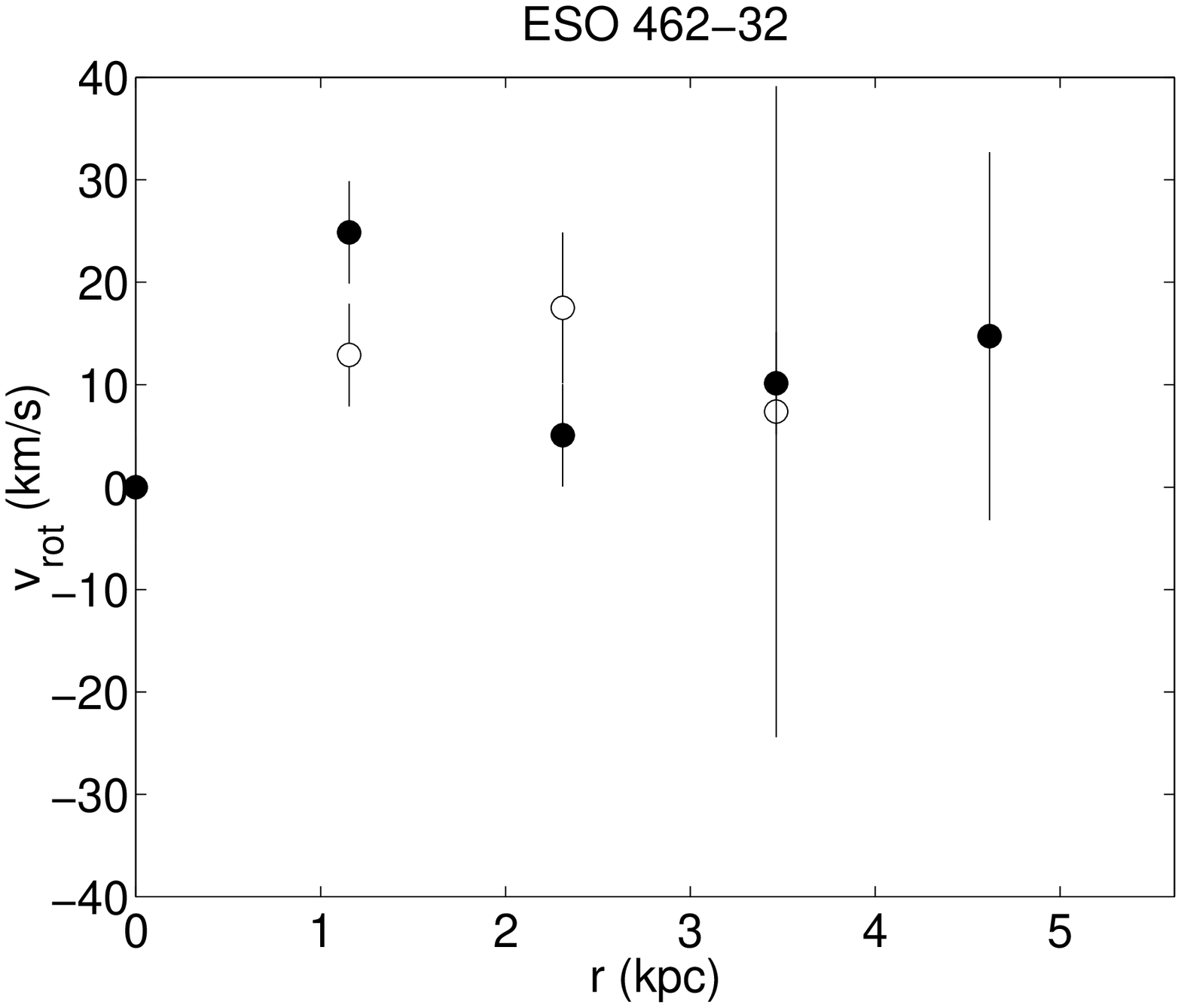}\includegraphics{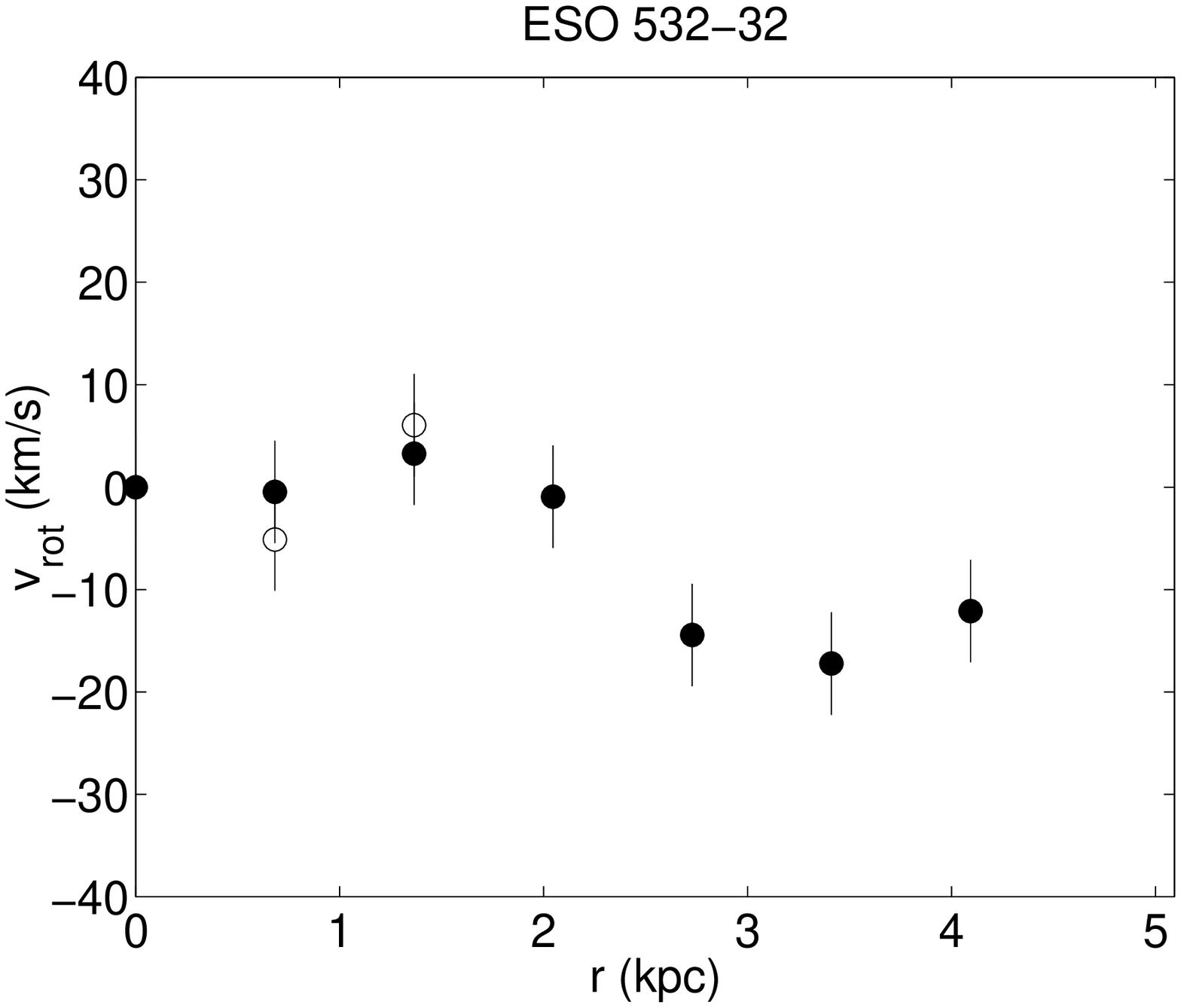}}
\includegraphics[scale=0.34]{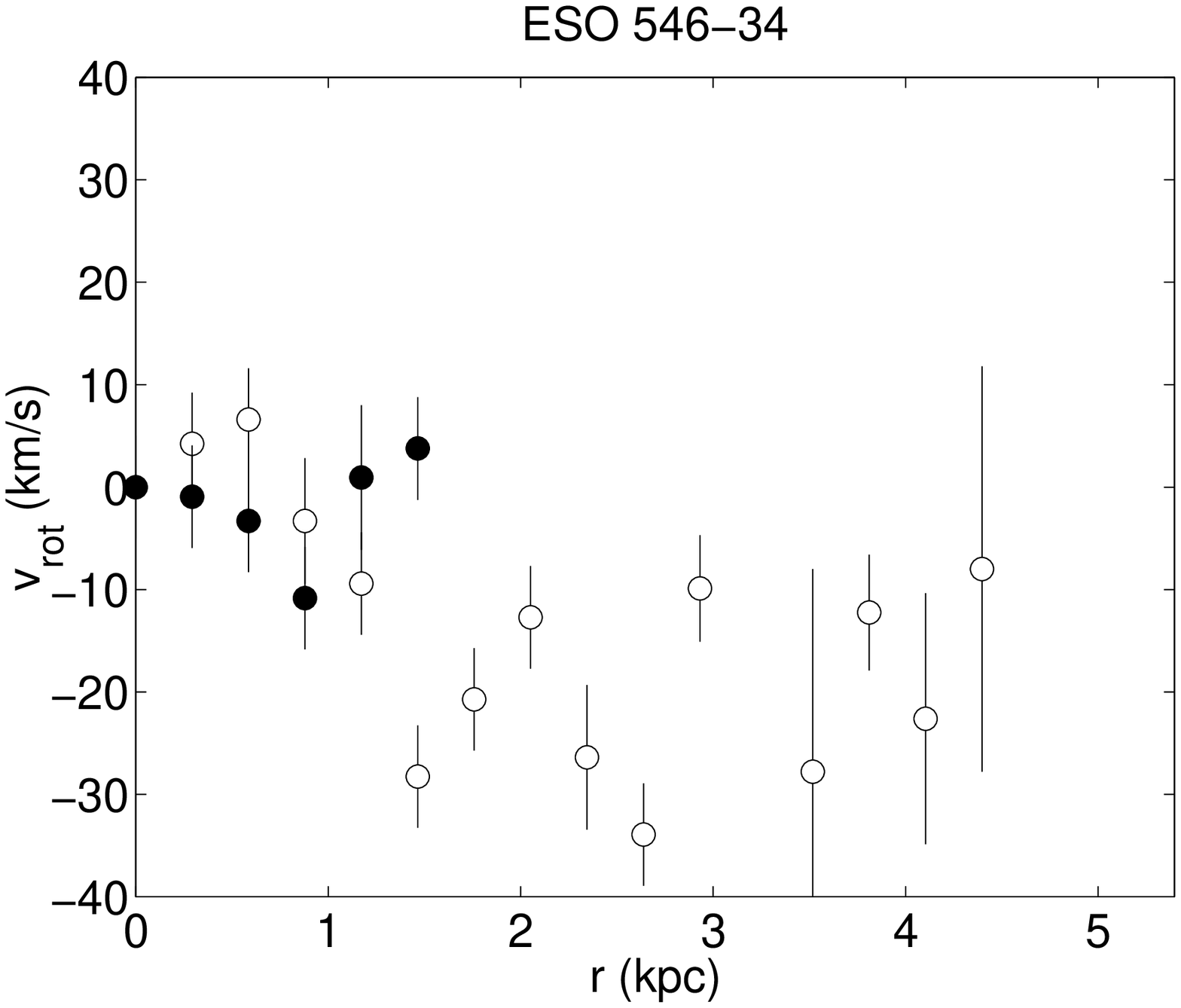}\includegraphics[scale=0.34]{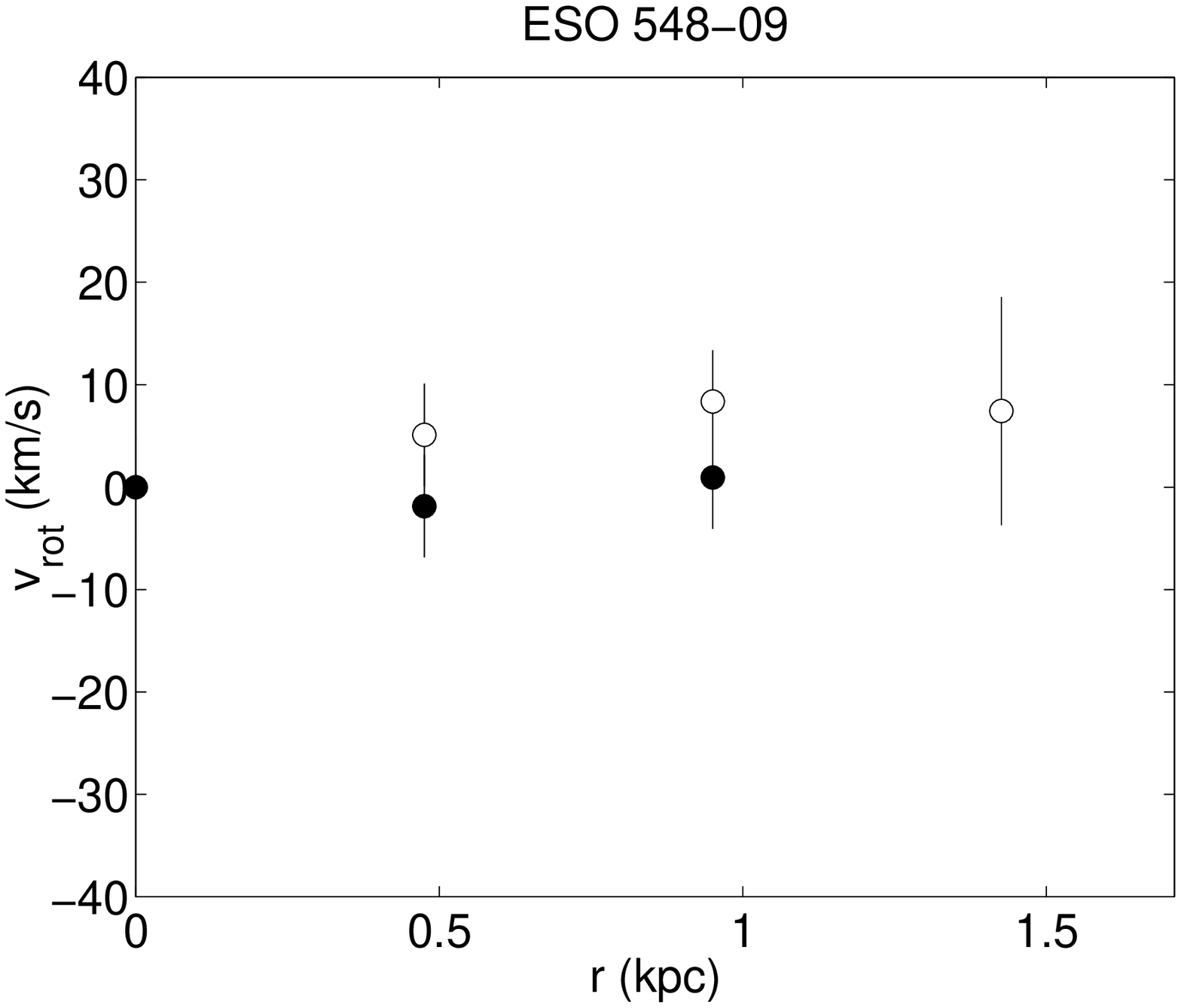}
\caption[]{The minor-axis velocity profiles of the target galaxies. Black and white filled circles represent data from opposing sides of the centre. The rotation curves have been folded spatially, but not in velocity. The radial axis has  been corrected for inclination.}
\label{fig_RC_minor}
\end{figure*}

Both the raw and the symmetrised, spatially averaged major-axis rotation curves are displayed in Fig.~\ref{fig_RC_major}. The corresponding minor-axis velocity profiles are shown in Fig.~\ref{fig_RC_minor}. The latter have been folded along the spatial axis only (not in velocity). Linear distances along the radial axis of the minor-axis velocity profiles have been corrected for inclination assuming an infinitely thin disk. Since these corrections are very large at high inclinations, the linear scale is admittedly highly uncertain, and in many cases in poor agreement with the scale of the major axis rotation curves.

A component of minor-axis rotation, as that found by Hayashi et al. (\cite{Hayashi et al. b}) in their simulation of a disk rotating in a triaxial CDM halo, would in Fig.~\ref{fig_RC_minor} turn up as velocities of opposite signs for the two sides. In no case is there any clear-cut evidence for this effect among our target objects.

In this paper, we make no detailed attempts to separate the different contributions (e.g. stellar disk, HI disk, dark halo) to the observed rotation curves, as HI rotation curves are currently lacking for these galaxies. While interesting constraints on the spatial distributions and relative masses of the different components may sometimes be inferred even without HI data (e.g. Borriello \& Salucci \cite{Borriello & Salucci}), the disc-halo degeneracies (e.g. van Albada et al. \cite{van Albada et al.}) in the region covered by optical rotation curves are often severe. An attempt to perform a mass decomposition of these galaxies will however be presented in a future paper (Mattsson et al., in preparation).

\subsection{Notes on individual galaxies}
Because of the high resolution of H$\alpha$ rotation curves and the typically patchy distribution of star-forming regions in LSBGs, irregularities of varying severity are present in all rotation curves presented here (although probably not more so than in previous investigations -- see e.g. de Blok \& Bosma \cite{de Blok & Bosma}; Swaters et al. \cite{Swaters et al.}). In a couple of cases, the rotation curves are so disturbed that there is little point in trying to uncover the underlying mass distribution from the observed kinematics. 

{\bf ESO 031-13:} The rotation curve is regular, except for at small wiggles at 3--6 kpc from the centre. The minor axis velocity profile shows no signs of significant non-circular motions.

{\bf ESO 146-14:} The rotation curve is full of wiggles and asymmetries. Indications of this were already present in the lower-resolution rotation curve of Bergvall \& R\"onnback (\cite{Bergvall & Rönnback}). Because of these small-scale irregularitites, a symmetrised, average rotation curve (admittedly highly uncertain) is constructed from very large radial bins. No minor-axis data is available (see Sect. 2.2).

{\bf ESO 462-32:} The rotation curve shows signs of non-circular motions in the centre, but is otherwise regular. The minor axis kinematics are unusual, with redshifted features on both sides of the centre. 

{\bf ESO 532-32:} While the major-axis rotation curve looks reasonably regular, the minor-axis velocity field does indicate the presence of non-circular motions at large distances from the centre.

{\bf ESO 546-34:} The major-axis velocity profile looks extremely strange with very little net rotation. The minor-axis data also indicates strong non-circular velocities. Although the outer isophotes of this galaxy appear fairly regular, there is an indication of a warp-like feature in the centre (see Fig.~\ref{Thumbnails}), reminiscent of two disks partly overlapping each other. It therefore seems likely that this is really two galaxies in a late stage of merging. Since no reliable mass profile can be derived from these data, no symmetrised and rebinned rotation curve is presented. 

{\bf ESO 548-09} The rotation curve displays significant wiggles (seen only on one side of the galaxy at a
 time) at around 1.3 and 2.3 kpc from the centre. The minor-axis velocity profile is consistent with no significant non-circular motions. 
\begin{figure*}[t]
\resizebox{\hsize}{!}{\includegraphics{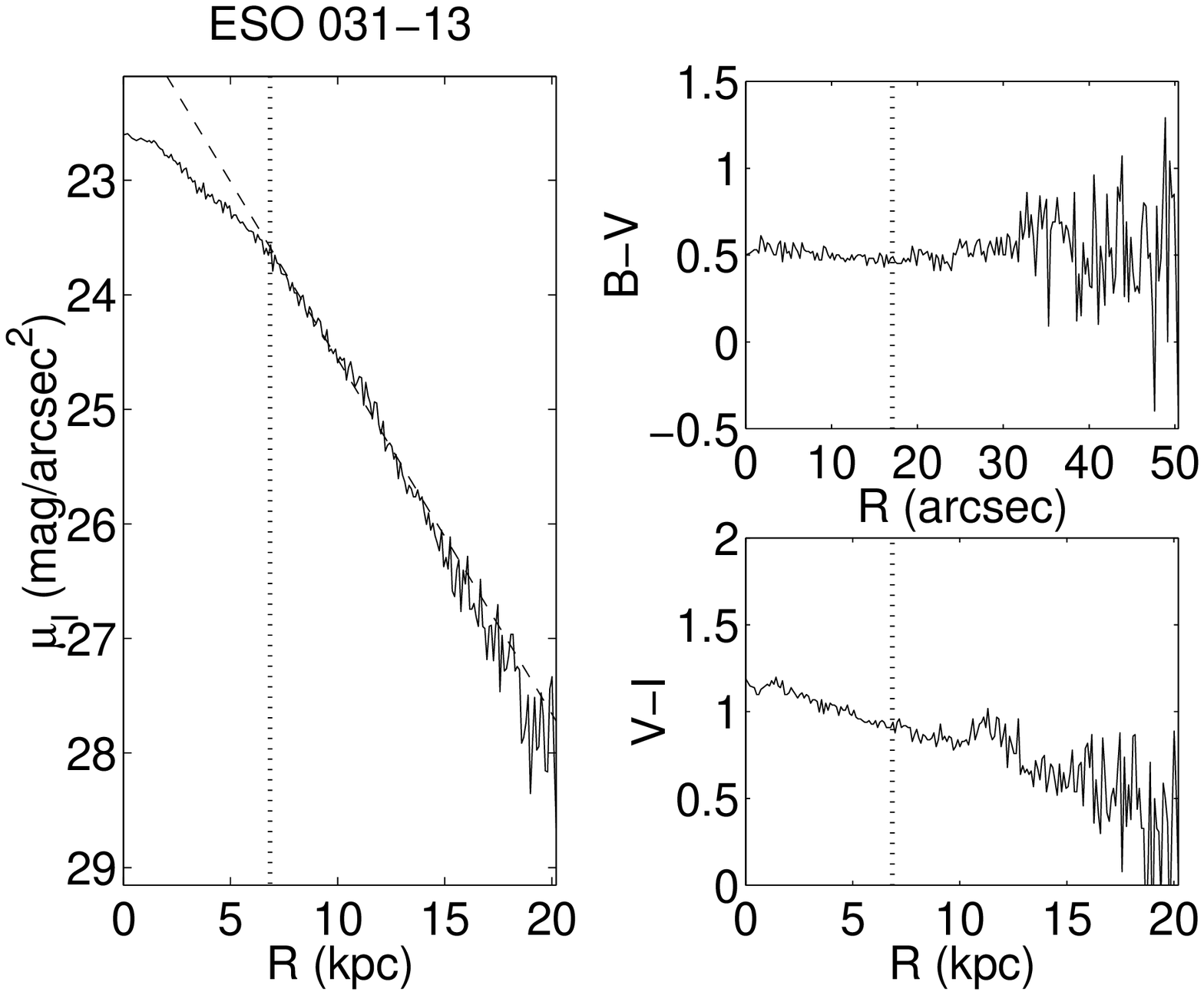}\includegraphics{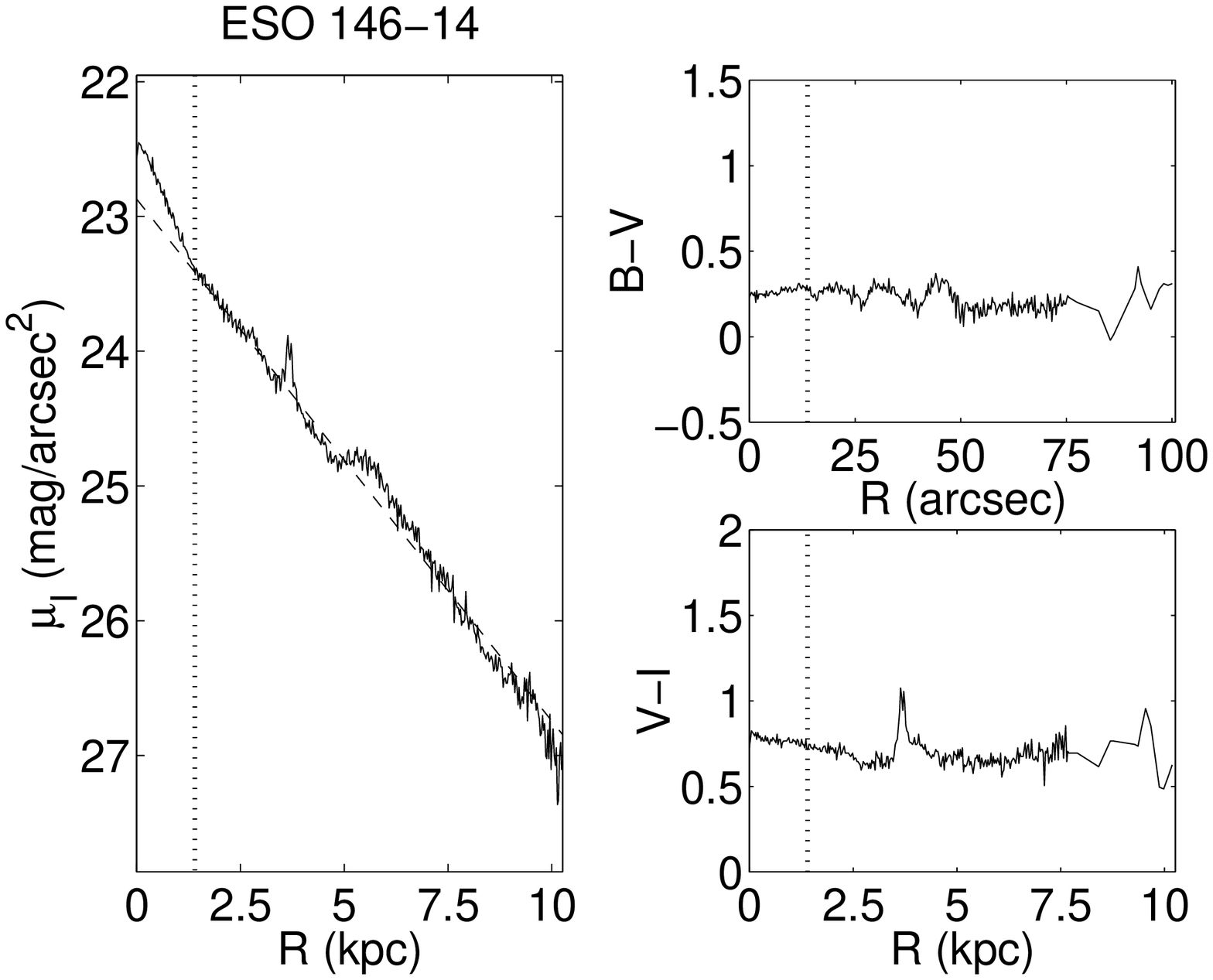}\includegraphics{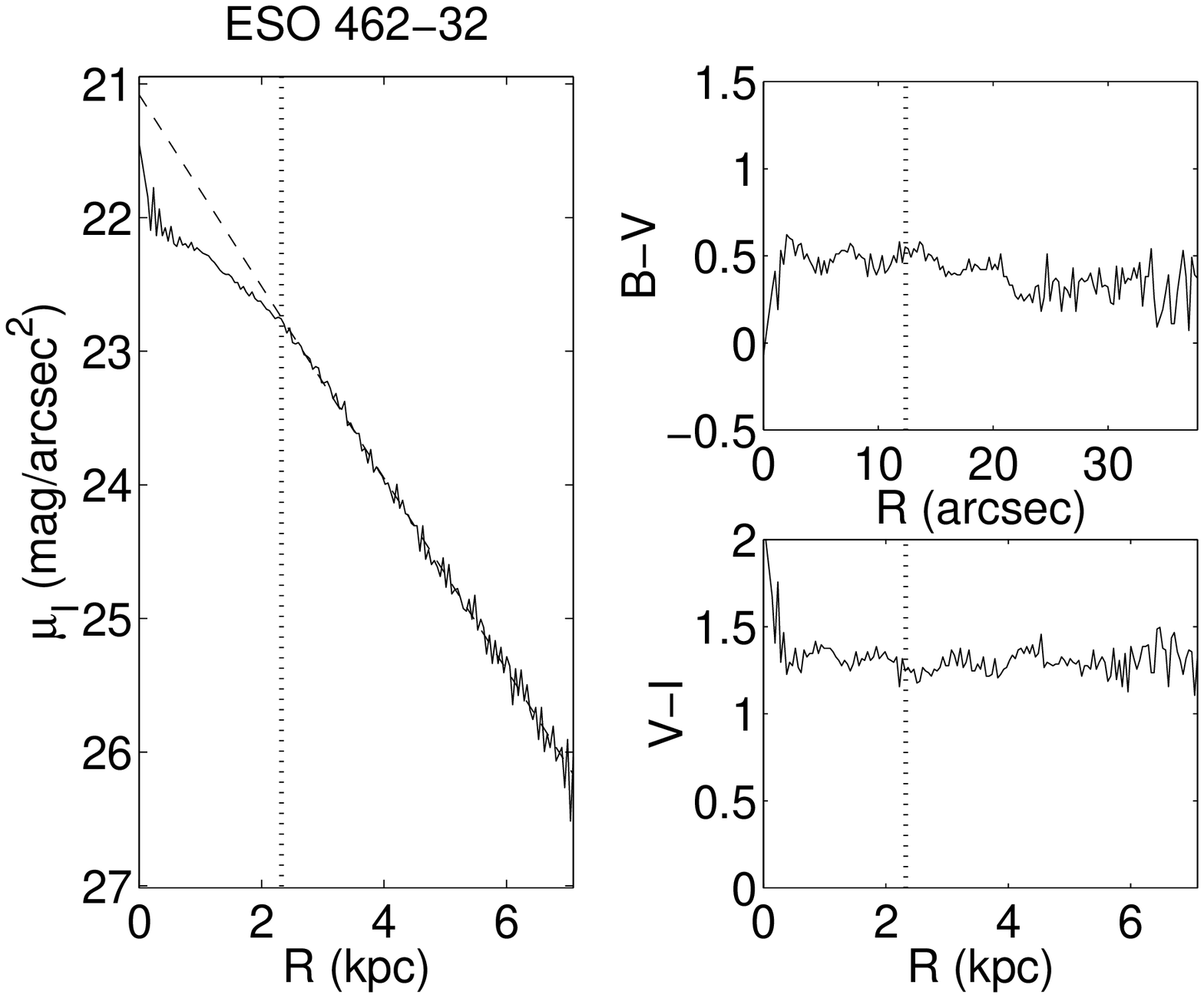}}
\resizebox{\hsize}{!}{\includegraphics{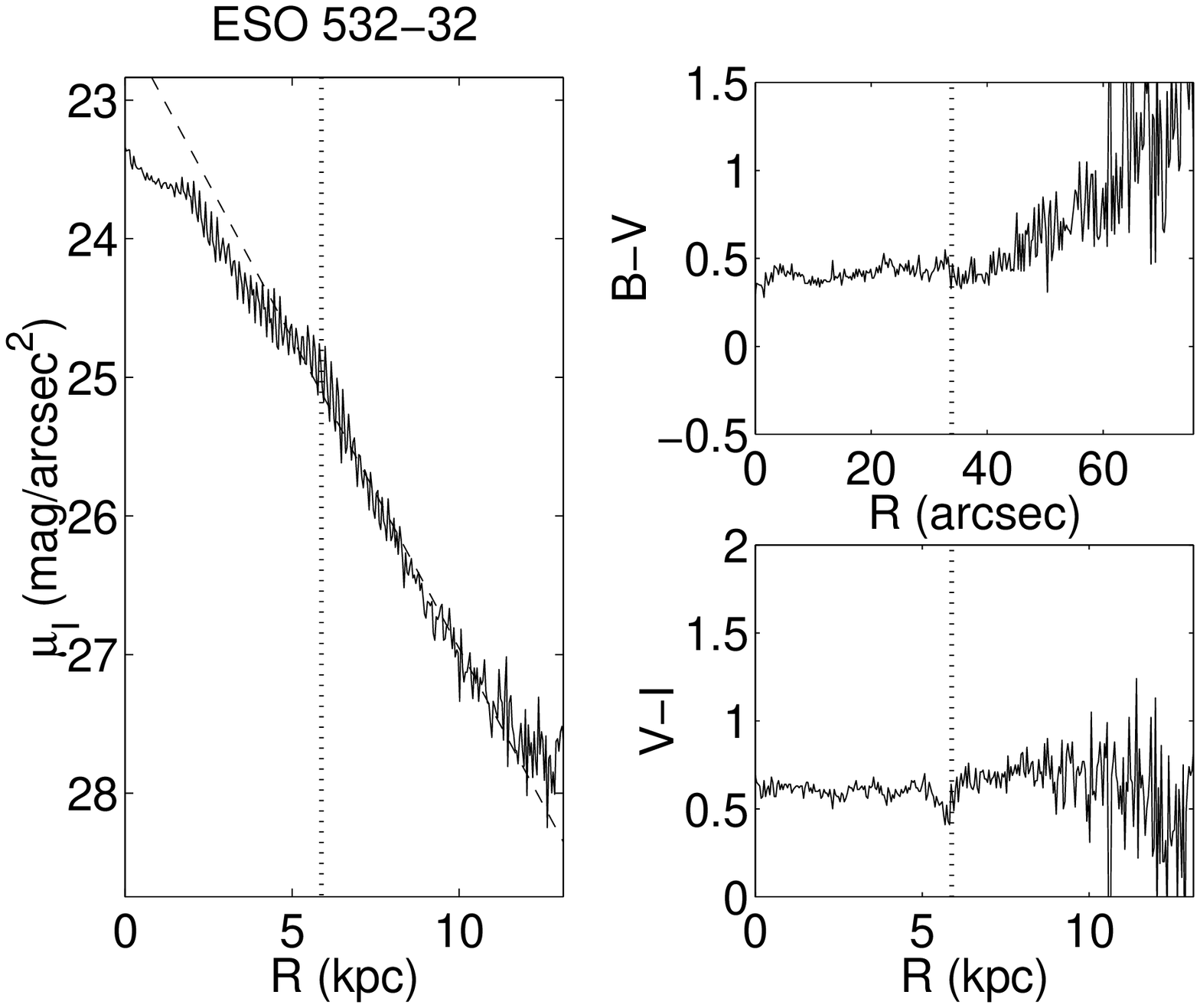}\includegraphics{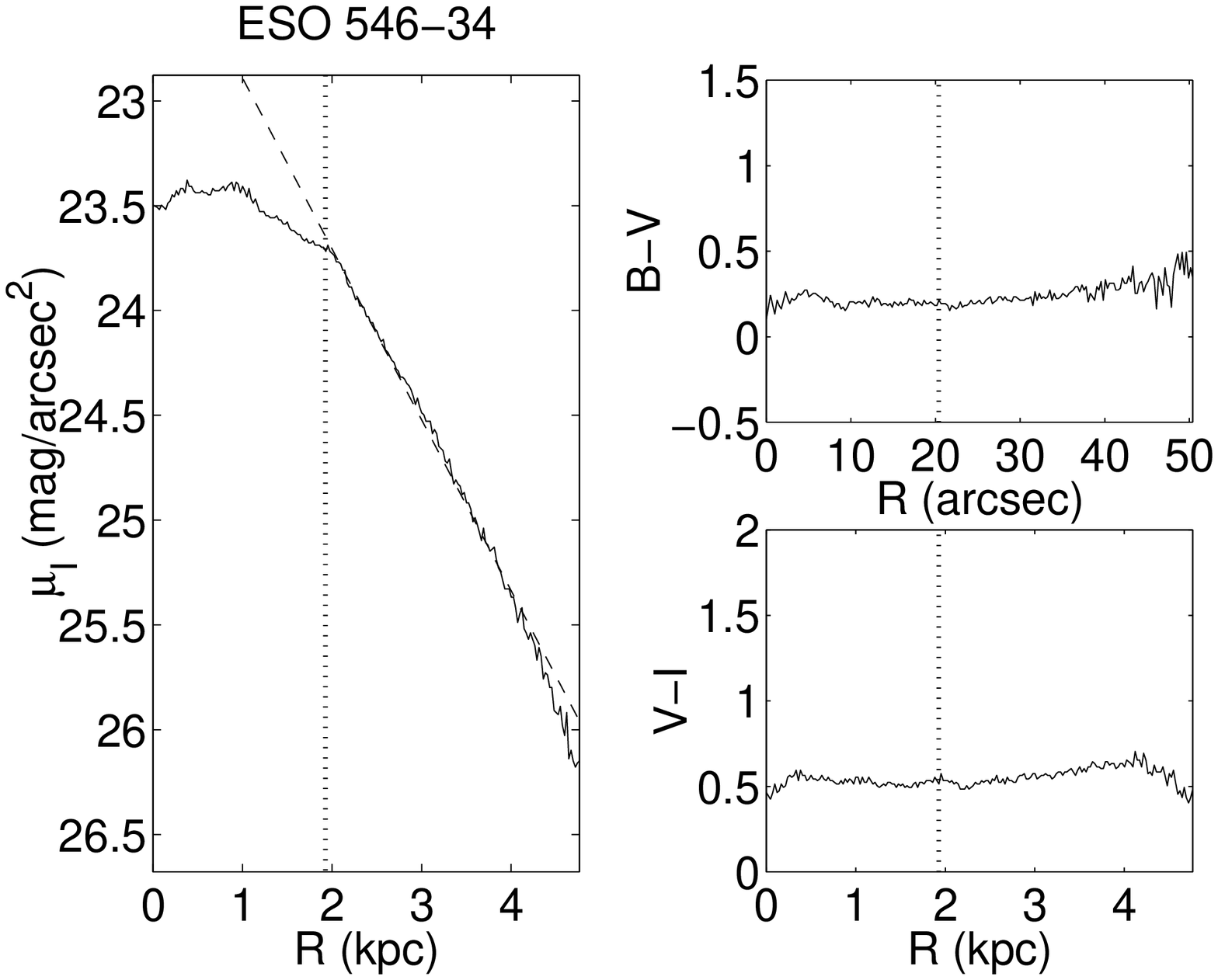}\includegraphics{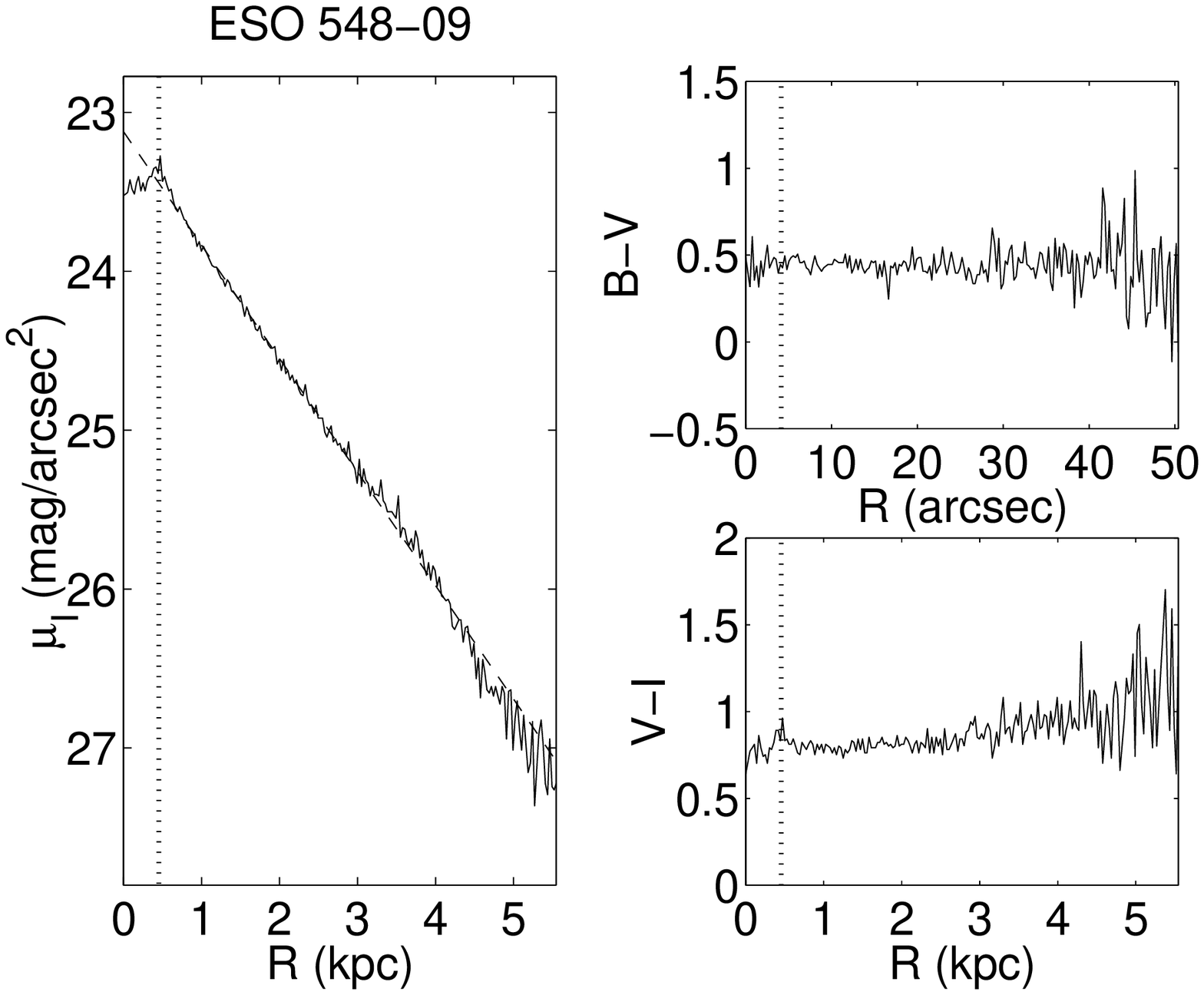}}
\caption[]{Radial $I$-band surface brightness profiles and $B-V$, $V-I$ colour profiles derived for the target galaxies. The surface brightness profiles have been corrected for inclination and Galactic extinction. The dashed lines represent the exponential disk profile determined from the outer isophotes. The radius at which the surface brightness profile starts to deviate substantially from an exponential disk is indicated by a vertical, dotted line. To allow a direct comparison between the linear and angular scales, arcseconds are used for the $B-V$ profile and kiloparsecs for $V-I$. The radial axes of the  graphs always extend to the same radius for each galaxy separately. Please note that the small-scale colour flucutations evident at large radii stem from decreasing signal-to-noise ratios in the outermost parts of the surface brightness profiles. To avoid cluttering, the error bars of the individual data points have not been included in this figure.}
\label{fig_BVI}
\end{figure*}
\section{Luminosity profiles} 
The $I$-band surface brightness profiles, depicted in Fig.~\ref{fig_BVI}, are derived by integrating the light over elliptical strips of constant orientation determined from the inclinations and position angles of the outer isophotes.

All galaxies except ESO 146-14 display a central light depression relative to the expectations from an exponential disk profile fitted to the outer parts of the galaxy. The $B-V$ and $V-I$ colour profiles typically do not show any dramatic gradients in the region when the depression starts to develop, indicating that this feature is unlikely to be an effect of dust reddening. The only exception is ESO 031-13, where the $I$-band profile is significantly steeper in the centre than $B$ or $V$. This $V-I$ gradient is however visible outside the break in the surface brightness profile as well. Since there is no corresponding slope in $B-V$, this is more likely to be due to a gradient in stellar population properties rather than to dust. In general, the colours vary very little across the face of the disks.

The reason for the central depression in the surface brightness profile of LSBGs is not well-understood. These profiles, which are seen in many blue LSBGs (e.g. R\"onnback \& Bergvall \cite{Rönnback & Bergvall a}, Bell et al. \cite{Bell et al.}) differ from the Freeman Type II profiles (Freeman \cite{Freeman}) seen in high surface brightness galaxies (e.g. MacArthur et al. \cite{MacArthur et al.}) in that the decrease in surface brightness persists all the way to the centre. Since no significant central colour gradients are seen in the optical/near-IR data (Bergvall et al. \cite{Bergvall et al. a}), these features cannot in general be attributed to dust effects. If this central light depression is related to some feature in the dark halo density profile, one could possibly expect to see some corresponding feature in the rotation curve or density profile at the radius at which the surface brightness profile starts to deviate from the fitted exponential disk. The radius which corresponds to the break points in the surface brightness profile have been indicated by vertical, dotted lines in the major-axis rotation curves (Fig.~\ref{fig_RC_major}) and the density profiles derived from these (Fig.~\ref{fig_densprof}). We do however not detect any obvious kinematic features associated with these breaks, except possibly for ESO 462-32, where the break coincides with the first data point of the plateau in the rotation curve. 

\begin{figure}[t]
\resizebox{\hsize}{!}{\includegraphics{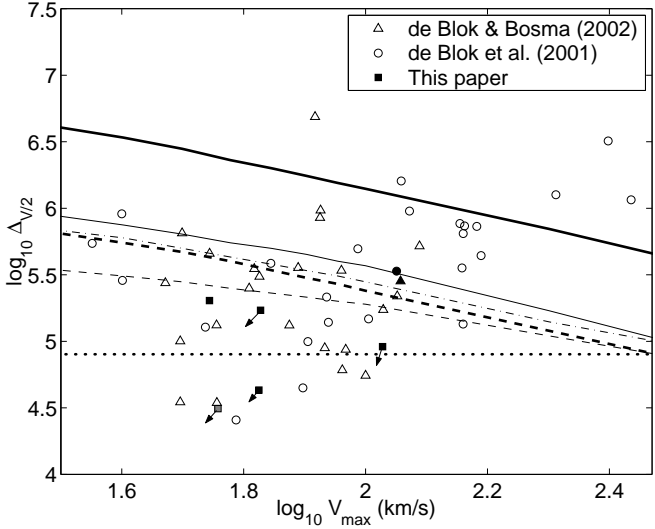}}
\caption[]{The central density parameter $\Delta_{V/2}$ as a function of the maximum circular velocity. Markers indicate LSBGs from de Blok et al. (\cite{de Blok et al. a}; circles), de Blok \& Bosma (\cite{de Blok & Bosma}; triangles)  and this paper (filled squares; ESO 146-14 in gray). 
The object F563-1, which appears in both the de Blok et al. and the de Blok \& Bosma samples, has been indicated by filled markers. The lines represent the predictions of various models: standard $\Lambda$CDM (Alam et al. \cite{Alam et al.}; thick solid), $\Lambda$CDM with a tilted power spectrum and slightly different cosmological parameters (Alam et al. \cite{Alam et al.}; thin solid), $\Lambda$CDM with additional 0.65 eV neutrinos (Zentner \& Bullock \cite{Zentner & Bullock}; thick dashed), $\Lambda$CDM with a running mass index and $n<1$ (Zentner \& Bullock \cite{Zentner & Bullock}; thin dashed), CDM with a dark energy equation of state parameter $w=-1.5$ (Kuhlen et al. \cite{Kuhlen et al.}; dash-dotted) and warm dark matter with a particle mass of 0.2 keV (Alam et al. \cite{Alam et al.}; dotted). Arrows indicate how our target galaxies would shift if a conservative correction for the baryonic disk was applied. For one object, ESO 548-09, this correction is too small to be seen. See main text for additional details.}
\label{centraldensfig}
\end{figure}
\begin{figure}[t]
\resizebox{\hsize}{!}{\includegraphics{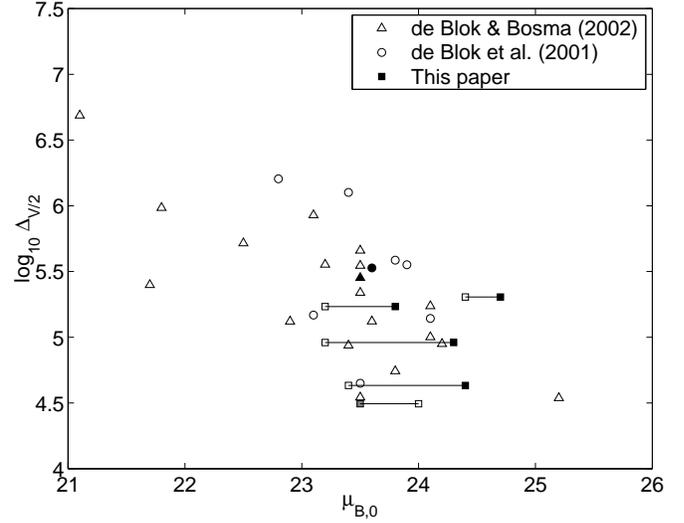}}
\caption[]{The dark halo central density parameter $\Delta_{V/2}$ as a function of central surface brightness in the $B$-band. Markers indicate LSBGs from de Blok et al. (\cite{de Blok et al. a}; circles), de Blok \& Bosma (\cite{de Blok & Bosma}; triangles) for which photometry is available. The object F563-1, which appears in both samples, has been indicated by filled markers. For our data, horizontal lines connect the true $\mu_{B,0}$ (filled squares; ESO 146-14 in gray) and that of an exponential disk extrapolated to the centre (open squares). See main text for additional details.}
\label{centraldensfig2}
\end{figure}

\section{The central density parameter}
The theoretically predicted dark halo properties can be compared with observations in several different ways. One method is to examine the absolute value of the dark matter density in the inner regions of the dark halo. In this context, Alam et al. (\cite{Alam et al.}) suggested the use of a simple, dimensionless central density parameter $\Delta_{V/2}$, defined as the mean dark matter density $\bar{\rho}$ (relative to the critical density of the Universe, $\rho_\mathrm{crit}$) within the radius $r_{V/2}$ where the galaxy rotation curve reaches half its maximum value $V_\mathrm{max}$:
\begin{equation}
\Delta_{V/2}=\frac{\bar{\rho}(r_{V/2})}{\rho_\mathrm{crit}}=\frac{1}{2}\left( \frac{V_\mathrm{max}}{H_0 r_{V/2}} \right) ^2.
\label{centraldenseq}
\end{equation}
Observationally, this quantity has the advantage that if the rotation curve rises at the outermost data point (which is common when only optical data is available), substituting $V_\mathrm{max}$ for the outermost (highest velocity) point of the rotation curve will -- for the range of likely density profiles -- result in an upper limit on $\Delta_{V/2}$ (see Alam et al. \cite{Alam et al.} for a discussion). Two other effects also contribute in the same direction. By applying (\ref{centraldenseq}) to the observed rotation curve without correcting for the rotational support provided by visible baryons (stars and gas), the central density of the dark halo will be overestimated. When comparing the observed and predicted dark halo $\Delta_{V/2}$ parameters, the observed value should also be lowered even further to correct for the halo contraction and density increase associated with baryonic cooling (e.g. Blumenthal et al. \cite{Blumenthal et al.}; Gnedin et al. \cite{Gnedin et al.}; Sellwood \& McGaugh \cite{Sellwood & McGaugh}), as current predictions for the dark halo $\Delta_{V/2}$ do not take this effect into account.

Despite the fact that $\Delta_{V/2}$ will be overestimated when derived directly from observed rotation curves, previous investigations have indicated that dwarf galaxies and LSBGs typically display central halo mass densities significantly {\it lower} than those predicted for CDM halos in a $\Lambda$CDM cosmology (e.g. Alam et al. \cite{Alam et al.}). The scatter in $\Delta_{V/2}$ among the observed halos is also higher than expected from simulations (e.g. Hayashi et al. \cite{Hayashi et al. a}). 

In Fig.~\ref{centraldensfig}, we plot $V_\mathrm{max}$ against $\Delta_{V/2}$ for  LSBGs from de Blok et al. (\cite{de Blok et al. a}; circles) and de Blok \& Bosma (\cite{de Blok & Bosma}; triangles) together with our targets (filled squares). In all cases, the outermost data points of the average rotation curves have been taken as proxies for $V_\mathrm{max}$. The corresponding $r_{V/2}$ have been estimated by linear interpolation between the data points of the rotation curve. Due to its disturbed rotation curve, the position of ESO 146-14 ($V_\mathrm{max}=57$ km/s and $r_\mathrm{V/2}=3.1$ kpc adopted) has been indicated by a gray marker. For this galaxy, the uncertainty mainly lies in the estimate of $r_\mathrm{V/2}$. Our $V_\mathrm{max}$ estimate is consistent with that derived by Bergvall \& R\"onnback (\cite{Bergvall & Rönnback}), and using the width of the HI profile at 20\% intensity instead of H$\alpha$ data gives $V_\mathrm{max}=74$ km/s (Mathewson \& Ford \cite{Mathewson & Ford}), which corresponds to only a minor shift of the data point unless the $r_\mathrm{V/2}$ estimate is severely off. Interestingly, our galaxies are all located in the lower part of the observed $\Delta_{V/2}$ distribution, where the conflict with theoretical predictions is the most severe. 

Included in Fig.~\ref{centraldensfig} is the predicted $V_\mathrm{max}$-$\Delta_{V/2}$ relation for a standard $\Lambda$CDM scenario (thick solid line; Alam et al. \cite{Alam et al.}) with a scale-free power spectrum of density fluctuations. Most of the data points are located at much lower halo densities than predicted by this model. The other lines represent various modifications of  standard $\Lambda$CDM suggested in the literature to improve the agreement with observations. Lower $\Delta_{V/2}$ may be achieved by tilting the power spectrum and adopting slightly non-standard values of the cosmological parameters $\Omega_\mathrm{M}$, $\Omega_\Lambda$, $\sigma_8$ and $H_0$ (Alam et al. \cite{Alam et al.}; thin solid), adding a small contribution of hot dark matter in the form of 0.65 eV neutrinos (Zentner \& Bullock \cite{Zentner & Bullock}; thick dashed), by assuming a running mass index at $n<1$ (Zentner \& Bullock \cite{Zentner & Bullock}; thin dashed line) or by assuming that the dark energy has an equation of state ($p=w\rho$) characterized by $w=-1.5$ (Kuhlen et al. \cite{Kuhlen et al.}; dash-dotted line). The only model which is in reasonable agreement with our observations is one of the warm dark models presented by Alam et al. (\cite{Alam et al.}; dotted line), where the dark matter is assumed to be a 0.2 keV fermion. It should be noted that the object ESO 120-021 from the sample of de Blok et al. (\cite{de Blok et al. a}), due to its very low $V_\mathrm{max}$, falls outside the range of the plotted model predictions in this diagram, and is therefore not included in the figure.
  
In this comparison, we have adopted the minimum disk hypothesis, i.e. that the dark matter halo completely dominates the density of these galaxies at all radii. If this (unrealistic) approximation is relaxed, our objects are expected to shift to even lower values of $\Delta_{V/2}$. To demonstrate this, an estimated minimum correction for the contribution from the baryonic disk to the observed rotation curve has been indicated by arrows in Fig.~\ref{centraldensfig}. These corrections are derived from $I$-band data, assuming an exponential stellar disk with scale lengths given in Table \ref{Obstable2}. In general, the existence of an empirical mass discrepancy-acceleration relation imposes strong constraints on the stellar population mass-to-light ratios of disc galaxies (e.g. McGaugh \cite{McGaugh a}). The colour-dependent mass-to-light ratios implied by this relation are, however, difficult to apply to our galaxies, since both models (Portinari et al \cite{Portinari et al.}) and empirical evidence (McGaugh \cite{McGaugh b}) indicates that the relation between colour and $M/L_\star$ goes noisy and non-linear for objects as blue as these. Here, we instead adopt the smallest $I$-band stellar population mass-to-light ratio ($M/L_\star=0.7$) found to be acceptable for the bluest LSBGs in the study of Zackrisson et al. (\cite{Zackrisson et al.}), under the assumption of a Salpeter initial mass function. While HI maps are lacking for the galaxies in our sample, the contribution from neutral gas to the rotation curves of LSBGs is usually found to be smaller than that of the stellar disk within the optical $r_\mathrm{V/2}$. (e.g. de Blok et al. a \cite{de Blok et al. a}; de Blok \& Bosma \cite{de Blok & Bosma}). We do however caution that the disk in many cases deviates substantially from the exponential form in the centre, and that this may affect the simple corrections applied here.

In Fig.~\ref{centraldensfig2}, we plot $\Delta_{V/2}$ against the $B$-band
central surface brightness levels $\mu_{B,0}$ for those objects in Fig.~\ref{centraldensfig} for which photometry is available. In the case of the de Blok et al. (\cite{de Blok et al. a}; circles) and de Blok \& Bosma (\cite{de Blok & Bosma}; triangles) data, the central surface brightness values correspond to that of an exponential disk extrapolated to the centre. For these objects, the $\mu_{B,0}$ values are in many cases based on $R$-band data, converted under the assumption of $B-R=0.9$ (following de Blok et al. \cite{de Blok et al. a}). For our data, both the true, central $\mu_{B,0}$ (filled squares) and that of the extrapolated exponential disk (empty squares) have been plotted. Despite a substantial scatter, the lack of faint $\mu_{B,0}$ objects at high $\Delta_{V/2}$ and bright $\mu_{B,0}$ objects at low $\Delta_{V/2}$ could possibly indicate the existence of a correlation between these two parameters.  While objects with widely different $V_\mathrm{max}$ have been included in this plot, binning in $V_\mathrm{max}$ does not seem to remove this correlation. There are at least two possible interpretations of a relation of this kind. The first is that dark halos on average have much lower central densities than the $\Lambda$CDM scenario predicts, and that objects at extremely low central surface brightness levels simply allow a much cleaner measurement of this density, as the correction for baryonic effects (which are usually assumed to increase the central densities) may be smaller for these objects. The other possibility is that extremely low surface brightness objects typically have dark halos with unusually low central densities, and that focusing on such objects when comparing to the {\it average} results from $\Lambda$CDM simulations therefore introduces an unfair bias. These different possibilities are further discussed in Sect. 7. The choice of true or disk $\mu_{B,0}$ in Fig.~\ref{centraldensfig2} is not crucial for our objects, although our observations do not seem very remarkable in terms of surface brightness when disk $\mu_{B,0}$ is plotted. In future investigations, it would be rewarding to explore how plots of central surface brightness versus $\Delta_{V/2}$ would look for objects with fixed $V_\mathrm{max}$ and rotation curves corrected for the contribution from visible baryons. 

\section{Slope of the central density profile}
Many attempts to test the predictions of the CDM model on the scales of galaxies have during recent years revolved around the so-called cusp/core problem. Whereas numerical simulations based on CDM have indicated that the central regions of dark halos should display a large increase in density -- a cusp (e.g. NFW) -- many observations have instead suggested the presence of a core of close to constant density (e.g.  Blais-Ouellette et al. \cite{Blais-Ouellette et al.}; de Blok \& Bosma \cite{de Blok & Bosma}; Gentile et al. \cite{Gentile et al. a}; Spekkens \& Giovanelli \cite{Spekkens et al.}). The original NFW dark halo density profile is given by:
\begin{equation}
\rho(R)=\frac{\rho_\mathrm{i}}{(R/R_\mathrm{S})(1+R/R_\mathrm{S})^2},
\end{equation}
where $R_\mathrm{S}$ represents a characteristic radius for the halo and $\rho_\mathrm{i}$ is related to the density of the Universe at the time of collapse.
In the very centre, this density profile is characterized by $\rho(r)\propto R^{\alpha}$, where $\alpha\approx -1$, i.e. a density cusp. The observations, on the other hand, often favour a value around $\alpha\approx0$, i.e. a constant-density core. From large samples of LSBGs and dwarf galaxies, both de Blok et al. (\cite{de Blok et al. b}) and Spekkens \& Giovanelli (\cite{Spekkens et al.}) find on average $\alpha\approx -0.2$. For this reason, cored density profiles have been advocated as superior dark halo models (e.g. Burkert \cite{Burkert}; de Blok et al. \cite{de Blok et al. b}). One commonly adopted such model is that of a pseudo-isothermal sphere, with a density profile given by:
\begin{equation}
\rho(R)=\frac{\rho_0}{1+\left(\frac{R}{R_\mathrm{C}}\right)^2} ,
\end{equation}
where $\rho_0$ is the central density of the halo and $R_\mathrm{C}$ the radius of the core for which this density is representative.

For several reasons, the slope of the central density profile may however not be the best test of CDM at the current time. On the theoretical side, there is still much controversy over the extent to which the NFW profile fitting formula really is appropriate in the very centre.  Whereas some authors have found central density slopes even steeper than the original NFW prediction (e.g. Fukushige \& Makino  \cite{Fukushige & Makino}; Reed et al. \cite{Reed et al.}), others argue in favour of a more shallow slope (e.g. Ricotti \cite{Ricotti}; Stoehr \cite{Stoehr}). On the observational side, a few studies have suggested that many measurements of the central density slope $\alpha$ may be biased by systematic effects so severe that observed cores may actually be consistent with intrinsic cusps (e.g. Swaters et al \cite{Swaters et al.}; Rhee et al. \cite{Rhee et al.}; Spekkens et al. \cite{Spekkens et al.}) -- but see e.g. de Blok et al. (\cite{de Blok et al. b}), de Blok (\cite{de Blok a}) and Gentile et al (\cite{Gentile et al. a}), for a different view.

\begin{figure*}[t]
\resizebox{\hsize}{!}{\includegraphics{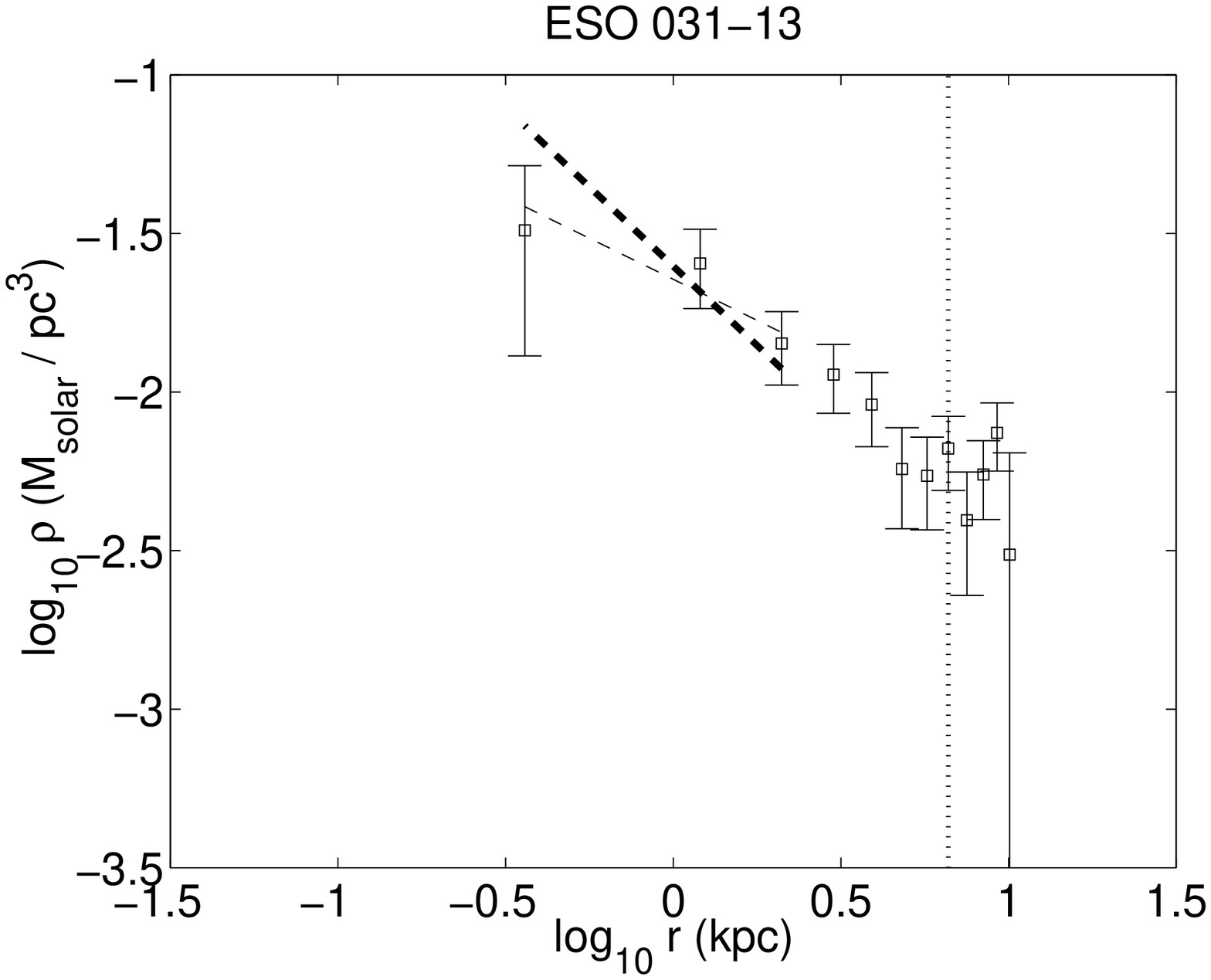}\includegraphics{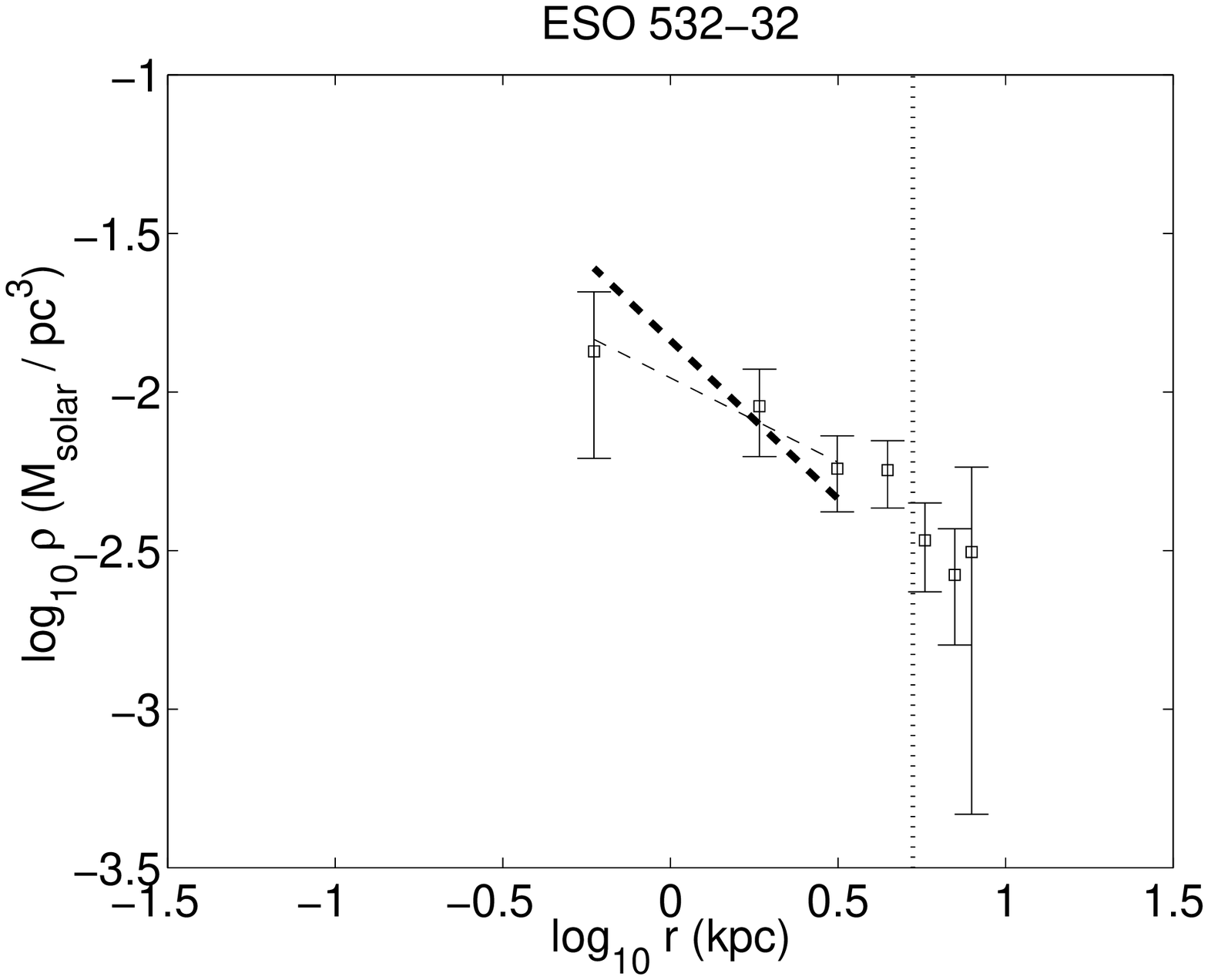}\includegraphics{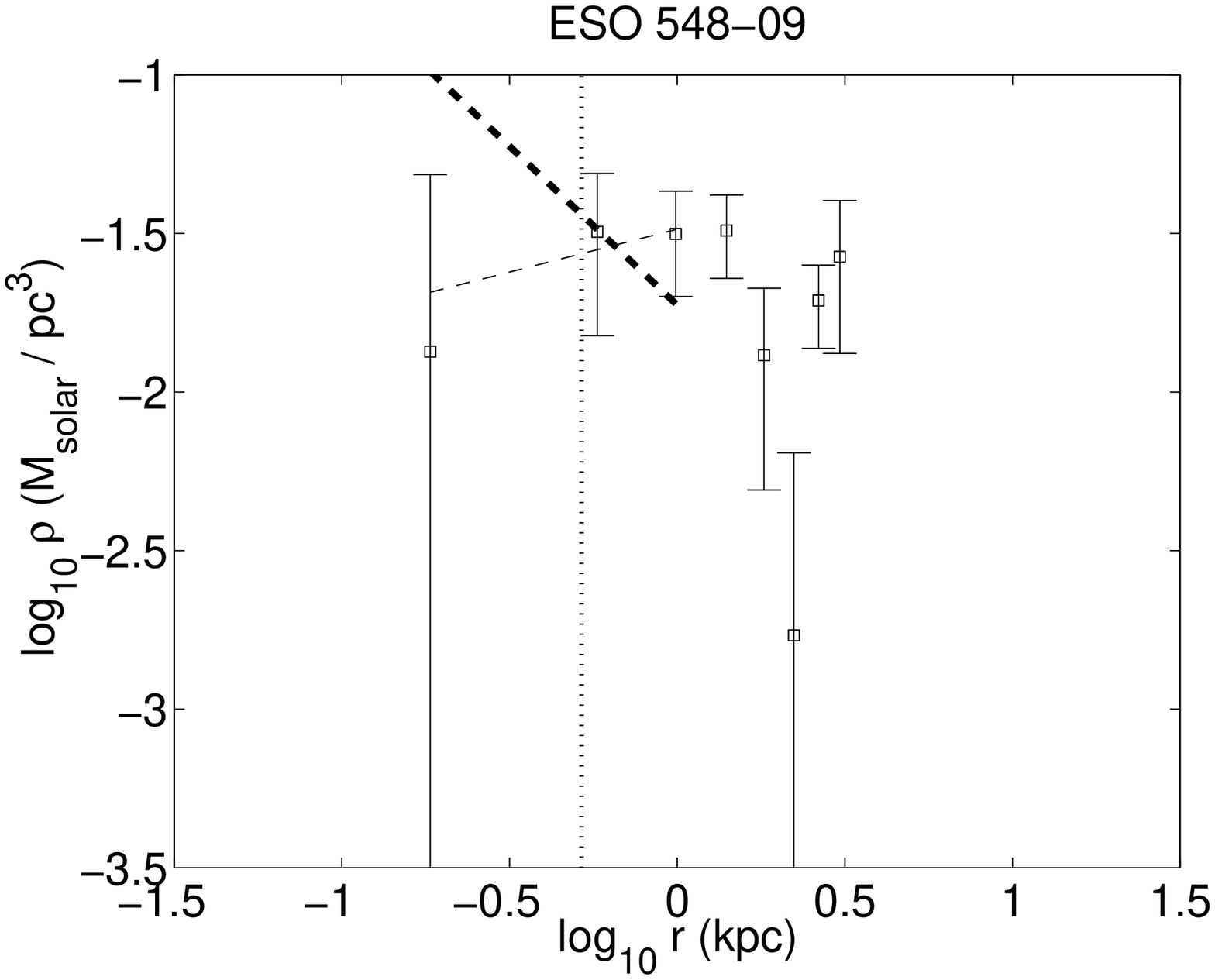}}
\caption[]{Density profiles derived under the assumption of a minimum disk and a spherical halo. Dashed lines indicate the power-law fitted to the innermost three data points of each profile. For comparison, the thick, dashed line indicates the forced fit of a cuspy density profile ($\alpha=-1$) to the same data. The vertical dotted line marks the break radius at which the surface brightness profile starts to deviate from the slope derived from the outer isophotes.}
\label{fig_densprof}
\end{figure*}

Despite these complications, it is nonetheless important to establish whether or not objects obeying our selection criteria deviate significantly from previous measurements of the central density profile. Under the assumption of a spherical matter distribution, the density profile can be calculated by direct inversion of the observed rotation curve $V(R)$:
\begin{equation}
\rho(R)=\frac{1}{4\pi G}\left( 2 \frac {V(R)\mathrm{d}V(R)}{R\mathrm{d}R} + \frac{V^2(R)}{R^2} \right).
\end{equation}
This procedure assumes that the baryons in the disk act merely as test particles and have a negligible impact on the dynamics (the minimum disk hypothesis). The resulting density profiles for the three objects (ESO 031-13, ESO 532-32 and ESO 548-09) with the least irregular velocity fields in the innermost region are displayed in Fig.~\ref{fig_densprof}.

\begin{figure}[t]
\resizebox{\hsize}{!}{\includegraphics{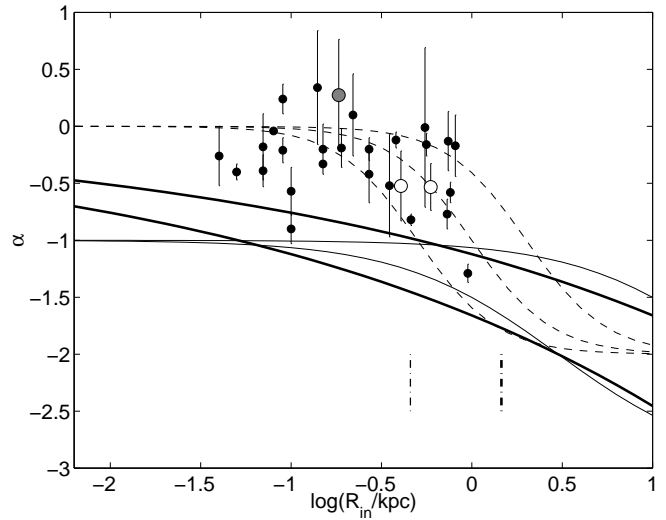}}
\caption[]{Central slope of the dark halo density profile ($\alpha$) vs. the radius of the innermost point in the rotation curve, $R_\mathrm{in}$. Markers indicate the measurements of $\alpha$ from our sample (large white/gray markers) and measurements from the LSBG sample of de Blok \& Bosma (\cite{de Blok & Bosma}; small black markers). Dashed lines indicate the density slopes of pseudo-isothermal halos with $R_\mathrm{c}=0.5$, 1 and 2 kpc (from left to right). Thin solid lines represent the density slopes of NFW halo profiles with $R_\mathrm{S}=3$ and 30 kpc (from bottom to top in the rightmost part of the plot). Thick solid lines represent the corresponding slopes from the updated Navarro et al. (\cite{Navarro et al. 04}) halo profile. The vertical thin  ($R_\mathrm{S}\approx 3$ kpc) and thick ($R_\mathrm{S}\approx 30$ kpc) dash-dotted lines indicate the innermost radii at which the Navarro et al. (\cite{Navarro et al. 04}) simulations are considered converged.}
\label{slopefig}
\end{figure}

To obtain the density slope $\alpha$, a power-law is fitted to the innermost three data points of the density profile. Following de Blok et al (\cite{de Blok et al. b}), the adopted error bars on $\alpha$ correspond to the maximum change in slope introduced by including one data point more or one data point less in the fit. In Fig.~\ref{slopefig}, the resulting central density slopes (large open markers) are compared to similar measurements (small black markers) by de Blok \& Bosma (\cite{de Blok & Bosma}) and various models for the dark halo density profile (lines). Following current convention (e.g. de Blok \& Bosma \cite{de Blok & Bosma}; Spekkens \& Giovanelli \cite{Spekkens et al.}) we plot $\alpha$ against $R_\mathrm{in}$ (representing the radius of the innermost point of the rotation curve), which gives an indication of the resolution of the observations. Since the rotation curve of ESO 548-09 becomes highly irregular at radii slightly outside that for which the density slope was derived, this object has been indicated by a gray marker. Within the error bars, our measurements are in reasonable agreement with the de Blok \& Bosma (\cite{de Blok & Bosma}) data and the average $\alpha\approx -0.2$ found by de Blok et al. (\cite{de Blok et al. b}) and Spekkens \& Giovanelli (\cite{Spekkens et al.}). They are furthermore in good agreement with the pseudo-isothermal sphere models (dashed lines, corresponding to core radii of $R_\mathrm{c}=0.5$, 1 and 2 kpc) advocated as models superior to the CDM predictions for the dark halos around galaxies. Due to the relatively large distances to the objects in our samples, our observations are not particularly impressive in terms of resolution. They do however indicate central density slopes substantially shallower than those predicted by the original NFW profile (thin solid lines, corresponding to $R_\mathrm{S}=3$ and 30 kpc -- suitable for halos in the dwarf-galaxy and large-galaxy mass range, respectively). Although the improved fitting formula (thick solid lines; same $R_\mathrm{S}$) presented in Navarro et al. (\cite{Navarro et al. 04}) predicts a more shallow density slope in the very centre, the discrepancy at the radii probed by our observations is not significantly smaller. In fact, at the smallest radii where the simulations of Navarro et al.  (\cite{Navarro et al. 04}) are considered converged (vertical thin and thick dash-dotted lines for halos with $R_\mathrm{S}\approx 3$ and 30 kpc, respectively) the agreement between the predictions ($\alpha\approx -1.4$ and $-1.2$, respectively) and the observations is {\it worse} than when the original NFW profile is used.

\section{Discussion}
The very low central mass concentrations (as measured by $\Delta_{V/2}$) of the halos around these blue LSBGs appear to be in serious conflict with the predictions of the halos formed in the $\Lambda$CDM scenario. We furthermore find no sign of the steep ($\alpha\leq -1$) CDM central density cusps, which are still advocated by some authors (e.g. Fukushige \& Makino \cite{Fukushige & Makino}; Navarro et al. \cite{Navarro et al. 04}; Reed et al. \cite{Reed et al.}).

Since those of our galaxies for which the inner density profile has been estimated appear to be completely bulgeless, the lack of the high central concentrations predicted for CDM halos contradicts the mechanism suggested by Mo \& Mao (\cite{Mo & Mao}) for lowering the central CDM halo density, as this model suggests that the original CDM halo profile should still be valid for bulgeless galaxies. Although only optical surface brightness profiles are presented in this paper, Bergvall et al. (\cite{Bergvall et al. a}) have shown that the decrease in surface brightness towards the centre (compared to an exponential disk profile), which is here interpreted as the absence of a bulge, typically persists in the near-IR as well. Hence, it cannot easily be attributed to dust effects.

Before claiming a general failure of the CDM paradigm, a number of additional complications do however require consideration.

\subsection{Issues related to disk inclination}
The inclinations of the galaxies in our sample have been estimated to lie in the range $i=76$--83$^\circ$. Although selecting high inclination disks allows the study of galaxies at fainter surface brightness levels than would otherwise have been possible, this particular property of the galaxies studied also opens the possibility that conclusions drawn from the measured rotation curves may be affected by projection effects. Matthews \& Wood (\cite{Matthews & Wood}) modelled the effects of internal extinction and projection effects in the disks of LSBGs, concluding that rotation curves should be negligibly affected at inclinations of $i\lesssim 85^\circ$ even at high optical depths. Some effects could however be discerned even at $i=85^\circ$ in the case of a very steep rotation curve slope. Similar calculations were carried out by Baes et al. (\cite{Baes et al.}), showing that projection and dust effects are expected to be very small, although not undetectable, at $i\lesssim 85^\circ$. The optimal inclination for accurate rotation curve retrieval appears to be around $i\approx 60^\circ$. Swaters et al. (\cite{Swaters et al.}), on the other hand, simulated the various inclination-related systematic effects involved in measuring the central density profile, with the alarming result that measurements of the inner slope of the density profile could be significantly biased even in galaxies with $i\approx 60^\circ$. These results were later questioned by de Blok et al. (\cite{de Blok et al. b}), who found the observed inner slope distribution to be insensitive to whether $i\leq80^\circ$ or $i\leq85^\circ$ was adopted in the galaxy samples used. Rhee et al. (\cite{Rhee et al.}) nonetheless detected an inclination-related bias of the type envisioned by Swaters et al. in the de Blok et al. data. 

Since the line of sight traverses the most diverse orbits of the observable disk at small distances from the centre, projection effects are however expected to become smaller further out (e.g. Rhee et al. \cite{Rhee et al.}). The concentration parameter $c$ (often defined as the ratio between the virial radius and the characteristic halo radius of the NFW profile: $c\equiv R_\mathrm{vir}/R_\mathrm{S}$) in many ways provides a more robust test than the innermost slope of the density profile, since it depends on the rotation curve behaviour over a much larger radial interval. Figs. 8 \& 9 of Swaters et al. (\cite{Swaters et al.}) do indeed confirm that the inclination-dependent systematic effects should be smaller for $c$ than for the innermost slope of the density profile. The related central density parameter $\Delta_{V/2}$ used here also extracts information from much larger radii, and should be less affected by inclination-related effects. 

In summary, a certain degree of scepticism should be applied when interpreting the inner slope of the density profile measured in high-inclination galaxies. The patchy distribution of H$\alpha$ emission seen at the very faint surface brightness levels of the galaxies analyzed here could further augment this problem. To put the conclusions reached in this paper on a more robust footing, it would therefore be very useful to extend this study to a sample of galaxies selected with similar colour and surface brightness criteria but lower inclinations.

\subsection{Issues related to the use of long-slit spectroscopy}
Aside from the extinction and projection effects that complicate the interpretation of rotation curves of high-inclination disk galaxies, there are also other observational effects that could affect the dark halo density profiles inferred from long-slit observations.  Wide slits, bad seeing and slits not properly positioned along the major axis would all make the intrinsically cuspy density profiles appear more core-like, as demonstrated by the simulations of de Blok et al. (\cite{de Blok et al. b}) and Swaters et al. (\cite{Swaters et al.}). With good seeing (here $\leq 0.8$\arcsec) and narrow slits (1\arcsec), the first two effects are however predicted to be modest. While slit offsets can certainly be a problem for individual objects with complicated isophotes (e.g. ESO 546-34 in our sample), the typical position errors of our observations are estimated to be no more than 1\arcsec, whereas offset errors of 3--4\arcsec would be required for intrinsically cuspy halos to mimic cores. 

\subsection{Issues related to the shapes of dark matter halos}
The analysis presented in this paper is based on the assumption of a spherical dark halo, while the CDM scenario in fact predicts dark halos to be triaxial. In Hayashi et al. (\cite{Hayashi et al. b}), it was argued that the discrepancy between observed LSBG rotation curves and the predicted properties of CDM halos could be reduced by properly taking the complicated dynamical effects of a disk rotating in a triaxial dark halo potential into account. As demonstrated in Sect. 3, our minor-axis data do however not show any sign of the kinematical signature derived by them for the elliptical disk that formed in their simulation. This is consistent with the results from other studies which also suggest that the observed non-circular motions in disks are insufficient to mask central density cusps (Gentile et al. \cite{Gentile et al. b}) and that the ellipticities of galactic disks in general are  small (see Combes \cite{Combes1} and references therein). Even so, this does not rule out the possibility that the apparent discrepancy between our results and the CDM predictions may become smaller once the approximation of a spherical halo is relaxed. As Hayashi et al. (\cite{Hayashi et al. b}) present only a single viewing angle from a single simulation, for which a particular orientation of the disk inside the triaxial dark halo was assumed, it is not at all clear how generic their results really are.

Recently, Dutton et al. (\cite{Dutton et al.}) investigated the effect of relaxing the assumption of spherical halos in favour of axially symmetric ellipsoids when interpreting rotation curve data, and found the best-fit halo concentration parameter to decrease for a halo flattened in the direction perpendicular to the disk, and to increase for a halo elongated in the same direction. As previous attempts to fit NFW halos to LSBG rotation curve data have indicated that the average concentration parameter $c$ tends to be too low compared to the CDM predictions  (e.g. McGaugh et al \cite{McGaugh et al.}; de Blok et al. \cite{de Blok et al. b}), a flattened halo would hence increase the tension between CDM predictions and LSBG observations, whereas elongated halos would decrease it. Dissipationless CDM simulations (e.g. Jing \& Suto \cite{Jing & Suto}) predict most halos to be prolate, i.e. having two axes of similar size and a third axis that is longer -- which is incidentally the halo shape adopted in the Hayashi et al. simulation as well. However, the typical halo shape can be altered substantially by baryonic processes (e.g. Kazantzidis et al. \cite{Kazantzidis et al.}) and may even become oblate (e.g. Dubinski \cite{Dubinski}), i.e. having two axes of similar size and third axis which is shorter. Indeed, many observational investigations (see Sackett \cite{Sackett} for a review) suggest that the dark halos of galaxies truly are oblate. As LSBGs are often assumed to be associated with dark halos of unusually high spin (e.g. Boissier et al. \cite{Boissier et al.}), and simulations indicate a correlation between high spin and oblateness (Moore et al. \cite{Moore et al.}), it seems quite reasonable to assume that LSBGs are located inside oblate halos. As hydrodynamic simulations of disks forming in dark matter halos furthermore indicate that the rotational axis of the disk tends to align with the minor axis of the inner halo (Bailin et al. \cite{Bailin et al. b}), it also seems reasonable to assume that an oblate halo should correspond to the polar flattening scenario investigated by Dutton et al. (\cite{Dutton et al.}). This suggests that relaxing the assumption of a spherical halo would just strengthen the case against CDM even further. There is, however, at least one potential flaw in this argument. While simulations predict a relation between the angular momentum of a dark halo and its shape, they also predict a relation between angular momentum and the concentration parameter $c$ (Bailin et al. \cite{Bailin et al. a}), which could bias measurements of $c$ in LSBGs towards values lower than the cosmic average (as further discussed in Sect. 7.5). 

To summarize, it is very difficult to say anything conclusively about how relaxing the assumption of a spherical dark halo would affect the comparison between CDM predictions and observations of LSBGs. To settle these issues, both improved predictions for the shapes of CDM halos under the influence of baryonic processes and methods for analyzing the kinematic data in the framework of these non-spherical halos would be required.  

\subsection{Issues related to the possible influence of baryons}
The reason for using LSBGs to test CDM halo predictions is that the effects of baryons on the overall density profile are believed to be minimized in these systems. Because of the strong correlation between dynamical mass-to-light ratio of and central disk surface brightness, the faintest LSBGs would appear to be the best targets available. Although the presence of undetected disk components in LSBGs have been claimed (Fuchs \cite{Fuchs}), this does not challenge the fact that in LSBGs, the radius at which the dark matter halo dominates over the maximum disk by a factor of two is quite small (McGaugh \& de Blok \cite{McGaugh & de Blok a}), although with a substantial scatter. More worrisome is the claim of Graham (\cite{Graham}), that the McGaugh \& de Blok relation may arise from selection effects, and that {\it all} LSBGs need not be as dark-matter dominated as previously assumed. Indeed, Bizyaev \& Kajsin (\cite{Bizyaev & Kajsin}) find that for LSBGs with bulges, the ratio of dark to luminous matter is not substantially different from that of high surface brightness galaxies. If the inner regions of LSBGs are not as CDM dominated as previously believed, but instead dominated by dark baryons (e.g. Combes \cite{Combes2}), this opens the possibility that some baryonic process -- e.g. bar formation (Athanassoula \cite{Athanassoula}) -- may have altered the inner density profile in some complicated way not taken into account in the simple scenario of adiabatic contraction (Blumenthal et al. \cite{Blumenthal et al.}; Gnedin et al. \cite{Gnedin et al.}; Sellwood \& McGaugh \cite{Sellwood & McGaugh}), which is often advocated to argue that baryons should increase -- and not decrease -- the central density. The commonly adopted procedures for analysing velocity profiles of disk galaxies can also lead to underestimates of both the central density and the slope of the central density profile if undetected bars are present (Rhee et al. \cite{Rhee et al.}). These issues need to be investigated further. 

\subsection{Issues related to dark halo selection effects}
Finally, there is the issue of whether or not the dwarf galaxies and LSBGs which are currently used to test CDM predictions on galactic scales constitute a biased dark halo sample. If LSBGs are preferentially formed in the low-density tail of the halo distribution, then discrepancies of the type highlighted here may perhaps not pose any serious threat to the CDM scenario (e.g. Zentner \& Bullock \cite{Zentner & Bullock}; Jimenez et al. \cite{Jimenez et al.}; Bailin et al. \cite{Bailin et al. a}), since comparing the dark halo properties of the galaxies with the lowest surface brightness levels to predictions for {\it typical} CDM halos would then be quite unfair. A bias of this kind can for instance arise if the halos surrounding LSBGs preferentially have an unusually late formation time (e.g. Jimenez et al. \cite{Jimenez et al.}) or if LSBGs are associated with dark halos of unusually high spin (e.g. Boissier et al. \cite{Boissier et al.}). Due to the degeneracy between age and star formation history, proving that LSBGs are young systems is however not trivial (Zackrisson et al. \cite{Zackrisson et al.}). It is furthermore not obvious that the cosmic distribution of halo spin is consistent with the observed surface brightness distribution of galaxies. In the model of Boissier et al. (\cite{Boissier et al.}), only a minority of disk galaxies ($\approx 35\%$) become LSBGs, which does not appear to be consistent with current estimates of the size of the LSBG population. With the LSBG definition  adopted by Boissier et al. ($\mu_{0,B}
\gtrsim 22$ mag arcsec$^{-2}$), the surface brightness distribution of disk galaxies presented by O'Neil \& Bothun (\cite{O'Neil & Bothun}) would for instance imply that $\gtrsim 80$ \% are LSBGs. McGaugh \& de Blok (\cite{McGaugh & de Blok a}) furthermore argue that the notion that LSBG reside in halos of unusually low density would be difficult to reconcile with the observed Tully-Fisher relation. Clearly, more detailed studies of halo concentration bias as a way to reconcile LSBG observations with CDM halo predictions are urgently needed.

\section{Summary}
By studying the H$\alpha$ kinematics of a sample of LSBGs with unusually blue colours and very low surface brightness centres, we find that:
\begin{itemize} 
\item These galaxies all fall inside the region of the diagram of maximum velocity $V_\mathrm{max}$ vs. central density parameter $\Delta_{V/2}$, which poses the toughest challenge for current $\Lambda$CDM dark halo predictions. The predictions of a number of alternative dark matter and dark energy models are also shown to be in conflict with these data.
\item By combining our observations with data from the literature, we uncover a possible relation between central surface brightness $\mu_{B,0}$ and the halo central density parameter $\Delta_{V/2}$. We argue that such a relation may explain the previous result and suggest that the faintest LSBGs -- in general -- may provide the most discriminating tests of $\Lambda$CDM on galactic scales, unless these galaxies systematically tend to form in the low-density tail of the dark halo distribution.
\item Our estimates of the slope of the innermost density profile in the centre of these galaxies are more reminiscent of constant density cores (consistent with previous investigations of other LSBGs and dwarf galaxies) than the steep density slopes predicted for CDM halos by either the original NFW fitting formula or the improved formula by Navarro et al. (\cite{Navarro et al. 04}).
\item Since most of the galaxies analysed are bulgeless, our inability to confirm the CDM halo predictions disfavours the scenario suggested by Mo \& Mao (\cite{Mo & Mao}), in which the halo density profile is altered by feedback associated with a fast collapse phase, as this model predicts that dark halos with the original CDM halo profile should still be detectable in bulgeless galaxies.
\item Our minor-axis data show no resemblance to the kinematical signature found by Hayashi et al. (\cite{Hayashi et al. b}) in their simulation of a disk rotating in a triaxial CDM halo. 
\item The inner radius at which the surface brightness profile starts to deviate from an exponential disk derived from the outer isophotes does not seem to correlate with any feature in the rotation curve or the estimated, spherically averaged density profile. The origin of these features remains a mystery.  
\end{itemize}
\begin{acknowledgements}
The authors would like to than the referee, G. Gentile, for very useful comments on the manuscript. EZ acknowledges financial support from the Royal Swedish Academy of Sciences. This research has made use of the NASA/IPAC Extragalactic Database (NED) which is operated by the Jet Propulsion Laboratory, California Institute of Technology, under contract with the National Aeronautics and Space Administration.  
\end{acknowledgements}

\end{document}